\documentclass[english,11pt]{article}
\pdfoutput=1
\usepackage[a4paper,bottom=4.2cm,top=1cm,head=3cm,width=18cm,dvipdfm]{geometry}
\evensidemargin=\oddsidemargin

\usepackage[T1]{fontenc}
\usepackage[latin9]{inputenc}
\usepackage{amstext}
\usepackage{babel}
\usepackage{verbatim}
\usepackage{amsfonts}
\usepackage{mathrsfs}
\usepackage{amsmath}
\usepackage{amssymb}
\usepackage{bm}
\usepackage{extarrows}
\usepackage{slashed}
\usepackage{isodateo}
\usepackage{xcolor}
\usepackage{graphicx}
\usepackage[low-sup]{subdepth}
\usepackage{isodateo}
\usepackage{caption}
\usepackage{subcaption}
\usepackage[linktocpage=true]{hyperref}
\hypersetup{
colorlinks=true,%
linkcolor=[rgb]{0,0,1}, 
citecolor=[rgb]{0,0,1}
}

\hypersetup{
	colorlinks=true,
	linkcolor=blue,
	filecolor=blue,
	urlcolor=blue,
	citecolor=blue,
}

\numberwithin{equation}{section}

\makeatletter
\@ifundefined{textcolor}{}
{%
	\definecolor{BLACK}{gray}{0}
	\definecolor{WHITE}{gray}{1}
	\definecolor{RED}{rgb}{1,0,0}
	\definecolor{GREEN}{rgb}{0,1,0}
	\definecolor{BLUE}{rgb}{0,0,1}
	\definecolor{CYAN}{cmyk}{1,0,0,0}
	\definecolor{MAGENTA}{cmyk}{0,1,0,0}
	\definecolor{YELLOW}{cmyk}{0,0,1,0}
}

\newcommand{\red}{\color{red}}

\newcommand{\blu}{\color{blue}}

\newcommand{\fr}[2]{\mbox{$\frac{\,{#1}\,}{#2}$}}
\renewcommand{\rm}{\mathrm}
\def\bge{\begin{equation}}
	\def\ede{\end{equation}}
\def\bga{\begin{aligned}}
	\def\eda{\end{aligned}}
\newcommand{\beq}{\begin{equation}}
	\newcommand{\eeq}{\end{equation}}
\newcommand{\bq}{\begin{equation}}
	\newcommand{\eq}{\end{equation}}
\newcommand{\ba}{\begin{array}}
	\newcommand{\ea}{\end{array}}
\newcommand{\beqa}{\begin{eqnarray}}
	\newcommand{\eeqa}{\end{eqnarray}}
\newcommand{\beqs}{\begin{subequations}}
	\newcommand{\eeqs}{\end{subequations}}

\def\nn{\nonumber}

\def\dis{\displaystyle}

\def\({\left(}
\def\){\right)}

\def\End{\end{document}}

\def\d{\text{d}}
\def\ii{{\tt i}}

\def\al{\alpha}
\def\be{\beta}
\def\ga{\gamma}

\def\ep{\epsilon}

\def\ito{\!\rightarrow\!}

\def\OT{\widetilde{\mathcal{O}}}
\def\OO{{\mathcal{O}}}

\renewcommand{\rm}{\mathrm}
\def\bge{\begin{equation}}
\def\ede{\end{equation}}
\def\bga{\begin{aligned}}
\def\eda{\end{aligned}}

\def\nn{\nonumber}

\def\dis{\displaystyle}

\def\({\left(}
\def\){\right)}
\def\[{\left[}
\def\]{\right]}

\def\End{\end{document}}

\def\al{\alpha}
\def\be{\beta}

\def\lam{\lambda}

\def\ga{\gamma}

\def\ep{\epsilon}
\def\lam{\lambda}

\def\si{\Delta}

\def\di{\mathrm{d}}

\def\to{\rightarrow}

\def\ii{\mathrm{i}}

\def\cut{\Lambda}

\newcommand{\mL}{\mathcal{L}}
\newcommand{\mO}{\mathcal{O}}

\def\TT{\mathcal{T}}
\def\vs{\vspace*{1mm}}
\def\hs{\hspace*{0.3mm}}
\def\hsx{\hspace*{0.5mm}}
\def\hsm{\hspace*{-0.3mm}}
\def\hsmx{\hspace*{-0.5mm}}

\def\to{\rightarrow}
\def\ito{\!\rightarrow\!}

\newlength{\halfpagewidth}
\divide\halfpagewidth by 2

\def\si{\sigma}

\def\End{\end{document}}

\begin{document}
\thispagestyle{empty}


\begin{center}
	{\Large\bf Probing CP-Violating Neutral Triple Gauge Couplings
		\\[2mm]
		at Electron-Positron Colliders}
	\vspace*{8mm}
	
	{\bf John Ellis}$\hs^{a,}$\footnote{Email: john.ellis@cern.ch},~~
	{\bf Hong-Jian He}$\hs^{b,}$\footnote{Email: hjhe@sjtu.edu.cn},~~
	{\bf Rui-Qing Xiao}$\hs^{c,}$\footnote{Email: xiaoruiqing@pku.edu.cn}
	
	
	\vspace*{3mm}
	{\small
		$^a$\,Tsung-Dao Lee Institute, Shanghai Jiao Tong University, Shanghai, China;\\
		Department of Physics, King's College London, Strand, London WC2R 2LS, UK;\\
		Theoretical Physics Department, CERN, CH-1211 Geneva 23, Switzerland
		\\[1mm]
		$^b$\,Tsung-Dao Lee Institute \& School of Physics and Astronomy, \\
		Shanghai Jiao Tong University, Shanghai, China;\\
		Department of Physics, Tsinghua University, Beijing, China;\\
		Center for High Energy Physics, Peking University, Beijing, China
		\\[1mm]
		$^c$\,School of Physics, Peking University, Beijing, China
	}
	
\end{center}

\vspace*{5mm}
\begin{abstract}
\baselineskip 16pt
\noindent
We study the CP-violating (CPV) neutral triple gauge couplings
(nTGCs) that can be realized via dimension-8 operators in the Standard Model
Effective Field Theory (SMEFT).\ We present a new formulation of the CPV nTGC form factors
that is compatible with spontaneous breaking of the electroweak gauge symmetry,
and show how these CPV form factors can be matched consistently with the corresponding dimension-8
CPV nTGC operators in the broken phase.\
We then study probes of the CPV nTGCs at future high-energy $e^+e^-$ colliders
with centre-of-mass energies $\sqrt{s\,}\!=\!(0.25,\hs 0.5,\hs 1,\hs 3,\hs 5)$\,TeV respectively,
demonstrating that the $e^{\mp}$ beam polarizations can help to improve
the sensitivities of probes of the nTGCs.\
We estimate that the sensitivity reaches for probing the new physics scales
of nTGCs can range from
${O}(\rm{TeV})$ at a 250\,GeV $e^+e^-$ collider
to ${O}(10\,\rm{TeV})$ at an $e^+e^-$ collider of energy $(3\hsmx -\hsmx 5)\hs$TeV,
and that the sensitivities to the nTGC form factors vary from
${O}(10^{-4})$ to ${O}(10^{-6}\hsm\!-\hsm\!10^{-8})$ for the $e^+e^-$ collision energy from 250\,GeV to $(3\!-\!5)\hs$TeV.
\\[8mm]
KCL-PH-TH/2025-01, CERN-TH-2025-002
\\[3mm]
Science China (Phys.\,Mech.\,Astron.) 68 (2025) 12, 121062 
{[{\hs}\href{https://doi.org/10.48550/arXiv.2504.13135}{arXiv:2504.13135}{\hs}].}
\\[3mm]
Selected as \,{\bf Editor's Focus}\, (Highlighted Article of Science China).
\\[1mm]
Journal's Research Highlights on this article:
\\
\hyperlink{}{http://engine.scichina.com/doi/10.1007/s11433-025-2812-x}
\\
\hyperlink{}{http://engine.scichina.com/doi/10.1007/s11433-025-2814-9}
\\
\hyperlink{}{http://engine.scichina.com/doi/10.1007/s11433-025-2813-4}
\end{abstract}


\newpage 
\tableofcontents

\baselineskip 17pt
\setcounter{page}{2}

\vspace*{5mm}
\section{\hspace*{-2.5mm}Introduction}
\label{sec:1}

Neutral triple gauge couplings (nTGCs) offer
a unique window to new physics
beyond the Standard Model (BSM): they are absent
in the Standard Model (SM) and cannot be generated
by dimension-6 interactions in the Standard Model
Effective Field Theory (SMEFT)\,\cite{SMEFT-Rev}, but provide direct
access to possible dimension-8 SMEFT interactions\,\cite{Degrande:2013kka}-\cite{nTGC-other}.\ 
Moreover, there are excellent prospects to 
search for nTGCs in both lepton-antilepton and 
hadron-hadron collisions using a variety of
final states involving photons and $Z$ bosons
that can decay into different identifiable
final states ($\ell^+ \ell^-$, $\nu \bar \nu$,
$q \bar q$), with distinctive kinematic
distributions that can be effectively discriminated from
the SM backgrounds.

\vs 

There have been many studies of nTGCs in the literature, 
including both the theoretical and experimental aspects, and using 
either a form factor framework\,\cite{Gounaris:1999kf} or 
the SMEFT formulation.\  In a recent series of works, we have studied
the present and prospective sensitivities of hadron-hadron colliders\,\cite{Ellis:2023ucy,Ellis:2022zdw} 
to nTGCs within the SMEFT framework, as well as the prospects of probing nTGCs
at the proposed electron-positron colliders\,\cite{Ellis:2025jgt,Liu:2024tcz,Ellis:2020ljj,Ellis:2019zex}.\ 
Along with these studies, we identified defects of the conventional form factor
parametrizations of nTGCs that had been applied previously by both the ATLAS and CMS
collaborations for their LHC experimental analyses:\ 
these were consistent with U(1)$_\rm{EM}^{}$ gauge symmetry but did not take
account of the full electroweak gauge symmetry  
SU(2)$_\rm{L}^{}\otimes\hs$U(1)$_\rm{Y}^{}$
of the SM that is incorporated in the SMEFT approach from the start.\ 
As we have shown\,\cite{Ellis:2023ucy}\cite{Ellis:2022zdw},
the conventional nTGC form factors based on the residual U(1)$_\rm{EM}^{}$ 
gauge symmetry alone are generally inconsistent 
and thus do not give reliable predictions.

In this work, we extend our previous analysis of the CP-conserving (CPC) nTGCs 
to include CP-violating (CPV) nTGCs in both the dimension-8 SMEFT formalism 
of nTGCs and the corresponding nTGC form factor formalism [compatible with spontaneous breaking of the SM electroweak gauge symmetry SU(2)$_\rm{L}\otimes\hs$U(1)$_\rm{Y}\hs$].\ 
We are motivated to study this
extension by the fact that the SM CP-violation as incorporated 
through the Kobayashi-Maskawa mechanism\,\cite{KM} is insufficient
to generate the observed baryon asymmetry of the Universe,
suggesting that there must be some BSM source of CP violation.\ 
This might appear at any scale beyond that of electroweak
symmetry breaking, and it could well appear at a scale
close to those already explored by the LHC experiments.\ 
Therefore it is important to scour all the CP-violating
signatures that are accessible to present and forthcoming
experiments.\ There have been relatively few studies of
experimental sensitivities to CPV nTGCs and, as
we show in this work, the conventional form factor formalism for the CPV nTGCs 
in the literature has serious defects, namely it is incompatible with 
spontaneous breaking of the full electroweak gauge symmetry of the SM. 

\vs 

The structure of this paper is organized as follows.\ In Section\,\ref{sec:CPVnTGCs}, 
we introduce the CPV dimension-8 SMEFT operators relevant for nTGCs 
and the corresponding CPV nTGC form factors that are compatible with 
spontaneous breaking of the SM electroweak gauge symmetry
SU(2)$_\rm{L}\otimes\hs$U(1)$_\rm{Y}$.\
Section\,\ref{sec:sensitivities} studies the sensitivities 
of probes of the CPV dimension-8 operators and the corresponding form factors 
that can be achieved in $e^+ e^-$ (or $\mu^+ \mu^-$) collisions 
at various collision energies up to 5\,TeV.\ 
The relevant cross sections are studied in
Subsection\,\ref{sec:cross-sections}, the correlations are analyzed in
Subsection\,\ref{sec:correlations}, and the potential for increasing the sensitivities
through a multivariable analysis is discussed in Subsection\,\ref{sec:MVA}.\ 
Our conclusions are summarized in Section\,\ref{sec:conx}.\  

\section{\hspace*{-2.5mm}CPV nTGCs and Form Factors from the SMEFT}
\label{sec:CPVnTGCs}
\label{sec:2}

In a series of previous works\,\cite{Ellis:2025jgt}-\cite{Ellis:2019zex},  
we have studied systematically the CP-conserving (CPC) dimension-8 SMEFT operators 
that generate the CPC nTGCs as well as their UV completion\,\cite{Ellis:2024omd}.\ 
We analyzed their contributions to helicity amplitudes 
and cross sections at both $e^+e^-$ colliders and hadron colliders.\  
In Refs.\,\cite{Ellis:2023ucy,Ellis:2022zdw},
we proposed an extended form factor formulation of the CPC nTGCs that is compatible with the 
full SU(2)$_{\rm L}^{}\otimes\hs$U(1)$_{\rm Y}^{}$ electroweak gauge symmetry of the SMEFT,
incorporating spontaneous breaking by the Higgs mechanism\,\cite{HiggsM},
and studied the prospective sensitivities of experimental
probes at both hadron-hadron colliders and electron-positron colliders.  

\vs 

In this Section, we analyze what CP-violating (CPV) nTGCs can be generated by the SMEFT operators 
of dimension-8 and match them to the corresponding CPV form factors that are compatible with
the electroweak gauge symmetry of SU(2)$_{\rm L}^{}\otimes\hs$U(1)$_{\rm Y}^{}$ in the broken phase.

\vspace*{1mm}

The general dimension-8 SMEFT Lagrangian can be expressed 
in the following form:
%
\beqa
\Delta\mathcal{L}(\text{dim-8})
\,=\, \sum_{j}^{}\!
\frac{\tilde{c}_j^{}}{\,\tilde{\cut}^4\,}\mathcal{O}_j^{}
\,=\, \sum_{j}^{}\!
\frac{\,\text{sign}(\tilde{c}_j^{})\,}{\,\cut_j^4\,}\mathcal{O}_j^{}
\,=\, \sum_{j}^{}\!
\frac{1}{\,[\cut_j^4]\,}\mathcal{O}_j^{}\,,
\label{cj}
\eeqa
where the $\,\tilde{c}_j^{}$ denote the dimensionless coefficients
of the dimension-8 operators $\mathcal{O}_j^{}$\,, which may take either sign and
sign$(\tilde{c}_j^{})\hsm\!=\hsm\!\pm$\,.\
For each operator $\mathcal{O}_j^{}$\,,
we denote its new physics cutoff scale as 
$\,\cut_j^{} \hsmx\equiv\hsm \tilde{\cut}/|\tilde{c}_j^{}|^{1/4}\,$,\
and for convenience we introduce the notation
$\,[\cut_j^4]\hsm\equiv\hsm\rm{sign}(\tilde{c}_j^{})\cut^4_j\hs$.

\vspace*{1mm}
	
We note that there are three dimension-8 CPV SMEFT operators including 
Higgs fields\,\cite{Degrande:2013kka}:
\beqs
\label{eq:dim8H} 
\begin{align}
\OT_{{B}W}^{} &\,=\,
		\ii\hs H^\dagger  {B}_{\mu\nu}W^{\mu\rho}
		\!\left\{D_{\!\rho}^{},D^\nu\right\}\! H+\text{h.c.} \hs,
\label{eq:obw}
\\
\OT_{{W}W}^{} &\,=\,
		\ii\hs H^\dagger  {W}_{\mu\nu}W^{\mu\rho}
		\!\left\{D_{\!\!\rho}^{},D^\nu\right\}\! H+\text{h.c.} \hs,
\label{eq:oww}
\\
\OT_{{B}B}^{} &\,=\,
		\ii\hs H^\dagger{B}_{\mu\nu}B^{\mu\rho}
		\!\left\{D_{\!\!\rho}^{},D^\nu\right\}\! H+\text{h.c.} \hs,
\label{eq:obb}
\end{align}
\eeqs
where $H$ denotes the Higgs doublet of the SM.\ 
From these, we can derive the following neutral triple gauge vertices (nTGVs):
\beqs
\label{eq:ZAV-dim8WB} 
\begin{align}
\hspace*{-5mm}
\Gamma_{Z\gamma \gamma^*}^{\alpha\beta\mu}
(q_1^{},q_2^{},q_3^{}) &=
e v^2q_3^2\!\(\!-\frac{s_W }{4c_W[\cut_{WW}^4]~} \!+\! \frac{1}{2[\cut_{WB}^4]} \!-\!\frac{c_W}{s_W[\cut_{BB}^4]}\!\)\hsm\!
\big(q_2^\alpha g^{\mu\beta}\hsm\!-\hsm q_2^\mu g^{\alpha\beta}\big)
\hs,
\\[1mm]
\hspace*{-5mm}
\Gamma_{Z\gamma Z^*}^{\alpha\beta\mu}
(q_1^{},q_2^{},q_3^{})
&=
e v^2(q_3^2\!-\!M_Z^2)\hsm\!\hsm\(\!\!-\frac{1}{\,4[\cut_{WW}^4]\,}
\!+\!\frac{c_W^2\!-\hsm s_W^2}{\,4c_Ws_W[\cut_{WB}^4]\,}
\!+\! \frac 1 {\,[\cut_{BB}^4]\,}\!\)\!\!
(q_2^\alpha g^{\mu\beta}\!\!-\hsm q_2^\mu g^{\alpha\beta})
\hs,
\end{align}
\eeqs
where $(s_W^{},\,c_W^{})\!=\!(\sin\theta_W^{},\,\cos\theta_W^{})$ 
and $\theta_W^{}$ is the weak mixing angle.\ 
Pure gauge operators containing only 
$B$ fields or only $W$ fields cannot contribute to the $Z\ga V^*$ vertex
via CPV nTGCs. 
However, the following two pure gauge operators that
contain both $B$ and $W$ fields can contribute to 
the CPV nTGCs of $Z\ga V^*$:
\\[-8mm]
\beqs
\label{eq:OG+G-}
\begin{align}
\label{eq:OG+}
g \OT_{G+}^{} &=\,	
{B}_{\!\mu\nu}^{}	 W^{a\mu\rho}
( D_\rho^{} D_\lambda^{} W^{a\nu\lambda} \!+\! D^\nu D^\lambda W^{a}_{\lambda\rho}) \hs,	
\\[1.5mm]
\label{eq:OG-}
g \OT_{G-}^{} &=\, 		
{B}_{\!\mu\nu}^{} W^{a\mu\rho}
( D_\rho^{} D_\lambda^{} W^{a\nu\lambda} \!-\! D^\nu D^\lambda W^{a}_{\lambda\rho}) \hs.
\end{align}
\eeqs
From these, we derive the following CPV neutral triple gauge vertices:
\beqs
\label{eq:ZAV-dim8G+-}
\begin{eqnarray}
\Gamma_{Z\gamma \ga^*(+)}^{\alpha\beta\mu}
(q_1^{},q_2^{},q_3^{})
\!\!&=&\!\!	-\frac{~v q_3^2s_W\,}{\,[\Lambda^4_+] M_Zc_W~}
(q_2^{\alpha}g^{\mu\beta}M_Z^2 \!-\hsm q_3^2q_2^\mu g^{\alpha\beta}) \hs,
\label{eq:ZAV-dim8G+A*}
\\
\Gamma_{Z\gamma Z^*(+)}^{\alpha\beta\mu}  
(q_1^{},q_2^{},q_3^{})
\!\!&=&\!\!	-\frac{~v (q_3^2-M_Z^2)\,}{\,[\Lambda^4_+] M_Z~}
(q_2^{\alpha}g^{\mu\beta}M_Z^2 \!-\hsm q_3^2q_2^\mu g^{\alpha\beta}) \hs,
\label{eq:ZAV-dim8G+Z*}
\\
\label{eq:ZAV-dim8G-A*}
\Gamma_{Z\gamma \ga^*(-)}^{\alpha\beta\mu}
(q_1^{},q_2^{},q_3^{})
\!\!&=&\!\!	-\frac{~vM_Z q_3^2s_W\,}{\,[\Lambda^4_-] c_W~}
(q_2^\alpha g^{\mu\beta}\hsm\!-\hsm q_2^\mu g^{\alpha\beta}) \hs,
\\
\label{eq:ZAV-dim8G-Z*}
\Gamma_{Z\gamma Z^*(-)}^{\alpha\beta\mu}(q_1^{},q_2^{},q_3^{})
\!\!&=&\!\!0 \hs,
\hspace*{16mm}
\end{eqnarray}
\eeqs
where the three gauge bosons are defined as outgoing.\ 
We note that
$\OT_{G+}^{}$ and $\OT_{G-}^{}$ share similar features with 
their CPC counterparts $\OO_{G+}^{}$ and 
$\OO_{G-}^{}$ \cite{Ellis:2022zdw}\cite{Ellis:2020ljj},\  
namely, neither $\OT_{G+}^{}$ nor $\OO_{G+}^{}$ contributes 
to the reaction $f\bar f \ito Z\ga$ with right-handed fermionic initial states, 
and both $\OT_{G-}^{}$	and $\OO_{G-}^{}$ contribute to the $Z\ga\ga^*$ 
coupling but not the $Z\ga Z^*$ coupling.\ 
We further note that the operator $\OT_{WW}^{}$ also does not 
contribute to $f\bar f \ito Z\ga$ when its initial states are right-handed fermions, 
because it contains only the SU(2)$_\rm{L}$ gauge fields $W^{\mu\nu}$ 
and the summed contributions from Eqs.\eqref{eq:ZAV-dim8G+A*}-\eqref{eq:ZAV-dim8G+Z*} 
cancel for the  right-handed fermionic initial states.\  
On the other hand, the operator $\OT_{BB}^{}$ does contribute
to the reaction $f\bar f \ito Z\ga$ with initial states being right-handed fermions,  
because it contains a pair of U(1)$_\rm{Y}$ gauge fields $B^{\mu\nu}$.

\vspace*{1mm}

We present next the form factor formalism for the CPV nTGCs.\ 
We note that the conventional CPV form factors respect the residual U(1)$_{\rm{EM}}^{}$ 
gauge symmetry, and are given as follows\,\cite{Gounaris:1999kf}:
\beq
\label{eq:FF0-nTGC}
\Gamma_{Z\gamma V^*}^{\alpha\beta\mu}
(q_1^{},q_2^{},q_3^{}) =
\frac{~e\hs (q_3^2\! -\hsm\! M_V^2)\,}{\,M_Z^{2}~}\!\!
\[ h_1^V\!(q_2^\alpha g^{\mu\beta}\!\!-\hsm q_2^\mu g^{\alpha\beta})
\!+\! \frac{h_2^V}{\,M_Z^2\,}\hs q_2^{\alpha}
(g^{\mu\beta}q_2^{}\!\cdot\hsm q_3^{} \hsm -\hsm q_2^\mu q_3^\beta)\hsm\]\!.
\eeq
This conventional CPV nTGC 
form factor formula is incompatible with
spontaneous breaking of the full electroweak gauge group 
SU(2)$_{\rm L}^{}\otimes\hs$U(1)$_{\rm Y}^{}$.\footnote{%
It was shown\,\cite{Ellis:2022zdw} that 
the conventional CPC nTGC form factor formulation\,\cite{Gounaris:1999kf} is also
incompatible with spontaneous breaking of the full electroweak gauge group.\
Going beyond \cite{Gounaris:1999kf}, we presented in Ref.\,\cite{Ellis:2022zdw} 
a formulation of CPC nTGC form factors that is compatible with 
the spontaneous electroweak gauge symmetry breaking.}\ 
In the following we first analyze this inconsistency and then present our new, 
consistent formulation for the CPV nTGC form factors.

\vspace*{1mm}

By direct power counting, 
we deduce the following leading energy dependences of 
the contributions of the CPV nTGC form factors $h_i^V\!$  
to the scattering amplitudes for
$\hs \TT[f\bar f\ito Z\ga]\hs$:
\beqs
\label{eq:FT8}  
\begin{align}
\label{eq:FT8-ZL}
\mathcal{T}_{(8)}^{\text{L}}
&= {O}(E^3) \!\times\hsm h_1^V + {O}(E^5) \!\times\hsm h_2^V ,
\hspace*{18mm}
\\[1mm]
\label{eq:FT8-ZT}
\mathcal{T}_{(8)}^{\text{T}}
&= {O}(E^2) \!\times\hsm h_1^V \hs,
\end{align}
\eeqs
where the amplitudes $\mathcal{T}_{(8)}^{\text{L}}$ and
$\mathcal{T}_{(8)}^{\text{T}}$
denote the final states having longitudinal $Z_L^{}$ and 
transverse $Z_T^{}$ bosons respectively.\ 
We note that for $\mathcal{T}_{(8)}^{\text{T}}$ with the on-shell transverse state $Z_T^{}$,
the $Z_T^{}$ polarization vector $\ep_{\al}^{}(Z_T^{})$ is orthogonal to the photon momentum
$q_2^{\al}$ (which is anti-parallel to the $Z$ momentum $q_1^{\al}\!=\!-q_2^{\al}$
in the center-of-mass frame), so this leads to 
$\ep(Z_T^{})\hsm\cdot\hsm q_2^{}\!=\!0\,$ and explains 
why the $h_2^V$ contribution vanishes in Eq.\eqref{eq:FT8-ZT}.\ 
However, we observe that the helicity amplitudes 
$\mathcal{T}_{(8)}^{\text{L}}$
contributed by the gauge-invariant dimension-8 nTGC
operators must obey 
the equivalence theorem (ET)\,\cite{ET}.\ 
For $E\!\gg\! M_Z^{}\hs$, the ET can be written as follows:
\\[-5mm]
\beq
\label{eq:ET}
\mathcal{T}_{(8)}^{}[Z_L^{},\ga_T^{}] \,=\,
\mathcal{T}_{(8)}^{}[-\ii\pi^0,\ga_T^{}] + B\,,
\eeq
where the longitudinal gauge boson $Z_L^{}$ absorbs 
the would-be Goldstone boson $\pi^0$   
through the Higgs mechanism, and the residual term 
$\,B\! =\!\mathcal{T}_{(8)}^{}[v^\mu Z_\mu^{},\ga_T^{}]\,$
is suppressed by a factor of
$\,v^\mu\!\equiv\!\epsilon_L^\mu\! -\!q_Z^\mu/M_Z^{}
\!=\! {O}(M_Z^{}/E_Z^{})$ \cite{ET}.\  
We note that the dimension-8 operators of Eq.\eqref{eq:dim8H} 
include the Higgs doublet and can contribute to the Goldstone amplitude 
$\mathcal{T}_{(8)}^{}[\pi^0,\ga_T^{}]$ at ${O}(E^3)$.\ 
On the other hand, the pure gauge operators of Eq.\eqref{eq:OG+G-} 
do not contain any Goldstone boson $\pi^0$ and thus give vanishing contribution 
to the Goldstone amplitude $\mathcal{T}_{(8)}^{}[\pi^0,\ga_T^{}]$
at tree level.\ But these pure gauge operators can contribute to 
the $B$-term of Eq.\eqref{eq:ET}, which is of ${O}(E^3)$.\  
Hence, consistency with the ET \eqref{eq:ET} requires that the ${O}(E^5)$ term on the right-hand-side (RHS) of 
Eq.\eqref{eq:FT8-ZL} must be cancelled such that its actual leading energy dependence
is ${O}(E^3)$.\ 

\vspace*{1mm}

Using this key observation, we can construct CPV nTGC form factors 
that are compatible with the full electroweak gauge group 
SU(2)$_{\rm L}^{}\otimes\hs$U(1)$_{\rm Y}^{}$ with spontaneous symmetry breaking.\
This requires that the $h_1^V$ term on the RHS of the conventional
CPV form factor formula \eqref{eq:FF0-nTGC} should be modified to
include a new form factor $h_6^V$, with the structure of  
$\,(h_1^V\!+ h_6^V q_3^2/M_Z^2)\hs$.\
Including $h_6$, we can consistently formulate the extended CPV nTGC form factor as follows:\footnote{%
We note that in the conventional formula \eqref{eq:FF0-nTGC}, 
the last term (containing $q_3^{\beta}$) vanishes in any physical application
because the momentum $q_3^{\beta}$ is contracted with the on-shell photon polarization vector 
$\ep_{\beta}^{}(\gamma)$.\ This contraction vanishes,
$q_3^{\beta}\ep_{\beta}^{}(\gamma)\!=\!0\hs$, because the photon polarization is transverse.\ 
This means that the last term of Eq.\eqref{eq:FF0-nTGC}
(containing $q_3^{\beta}$) can be discarded for all physical applications.\ 
Hence we drop this term in our new formulation
of the CPV nTGC form factors in Eq.\eqref{eq:ZAV*-FFnew}.}\ 
\beqs 
\label{eq:ZAV*-FFnew}
\begin{align}
\label{eq:ZAV*-FFnew-1}
\hspace{-6mm}
\Gamma_{Z\gamma V^*}^{\alpha\beta\mu}
&\!=
\frac{\,e\hs (q_3^2\hsm -\! M_V^2)\,}{\,M_Z^{2}~}\!\!
\[\!\Big(\hsm h_1^V\!\!+\!\hsmx\frac{\,h_6^Vq_3^2\,}{M_Z^2}\!\Big)\!
(q_2^\alpha g^{\mu\beta}\!\!-\hsm q_2^\mu g^{\alpha\beta})
\!+\!\hsm \frac{h_2^V}{\,2M_Z^2\,}\hs q_2^{\alpha}g^{\mu\beta}
\hsm (M_Z^2\!-\hsm q_3^2)\hsm\]
\\
\label{eq:ZAV*-FFnew-2}
\hspace{-6mm}
&\!=
\frac{\,e\hs (q_3^2\! -\! M_V^2)\,}{\,M_Z^{2}\,}\!\hsmx
\[\hsm h_1^V\hsmx (q_2^\alpha g^{\mu\beta}\!\!-\hsm q_2^\mu g^{\alpha\beta})\hsmx +\hsmx  
\frac{\,h_2^V\hsm M_Z^2\hs q_2^{\alpha}g^{\mu\beta}
\!\!-\!2h_6^Vq_3^2q_2^\mu g^{\alpha\beta}\,}{\,2M_Z^2\,}
\hsmx +\hsmx 
\frac{\,2h_6^V\!\!-\!h_2^V\,}{2M_Z^2}q_3^2q_2^\alpha g^{\mu\beta}\]
\hsm\!,
\end{align}
\eeqs 
where, inside the brackets, we see that at high energies the $h_1^V$ terms 
have leading energy (momentum) dependences that are $O(E^1)$,
whereas the $h_2^V$ and $h_6^V$ terms have leading energy dependences that are $O(E^3)$.   

However, we note that the brackets in Eq.\eqref{eq:ZAV*-FFnew-2} contain three terms, and that the middle part contains  
an $h_6^V$ term with leading energy dependence $O(E^3)$ that is proportional to 
$g^{\al\be}$ and will contract with the external $Z$ and $\ga$ polarization vectors:
$g^{\al\be}\ep_{\al}^{}\hsm (Z)\ep_{\be}^{}(\gamma)
\!=\hsm\ep^{\al}\hsm (Z)\ep_{\al}^{}(\ga)\hs$.\ 
In the case of the longitudinally-polarized external state $Z_L^{}$,
this contraction vanishes: $\ep^{\al}\hsm (Z_L^{})\ep_{\al}^{}(\ga_T^{})\!=\!0\hs$.\
This is because the spatial component of $\ep^{\al}\hsm (Z_L^{})$ is along the direction
of the 3-momentum $\vec{q}_1^{}$ and is thus orthogonal to the spatial part of the 
photon polarization vector, $\vec{q}_1^{}\!\cdot\vec{\ep}\hsx (\ga_T^{})\!=\!0\hs$,
where $\vec{q}_1^{}\!=\!-\vec{q}_2^{}$ holds in the center-of-mass frame of $Z\hs\ga\hs$.\  
From this we further deduce  $\ep^{\al}\hsm (Z_L^{})\ep_{\al}^{}(\ga_T^{})\!=\!0\,$
since the transverse photon polarization vector has vanishing time-component 
$\ep_{0}^{}(\ga_T^{})\!=\!0\hs$.\ 
This means that in Eq.\eqref{eq:ZAV*-FFnew-2} the $h_6^Vg^{\al\be}$ term in 
the middle part of the brackets has vanishing contribution to the amplitude
$\hs\TT[f\bar f\ito Z_L^{}\ga_T^{}]\hs$.
Finally, we note that the third term inside the brackets
is proportional to $(2h_6^V\!- h_2^V)$ 
and can contribute $O(E^5)$ to the amplitude
$\hs\TT[f\bar f\ito Z_L^{}\ga_T^{}]\hs$.\   

\vs  
 
Thus we deduce from Eq.\eqref{eq:ZAV*-FFnew-2} the
following leading energy dependences of the $h_i^V$ contributions
to the scattering amplitude $\hs\TT[f\bar f\ito Z\ga]\hs$:
\beqs 
\label{eq:FT8-ZLZT-h6}
\begin{align}
\label{eq:FT8-ZL-h6}
\mathcal{T}_{(8)}^{\text{L}}
&= {O}(E^3) \!\times\! h_1^V + {O}(E^3) \!\times\! h_2^V
+ {O}(E^5) \!\times\! (2h_6^V \hsm\!-\hsm h_2^V) \hs ,
\\[0.5mm]
\label{eq:FT8-ZT-h6}
\mathcal{T}_{(8)}^{\text{T}}
&= {O}(E^2) \!\times\! h_1^V + {O}(E^4) \!\times\! h_2^V
+ {O}(E^4) \!\times\! h_6^V \hs ,
\end{align}
\eeqs 
where the ${O}(E^5)$ terms originate from the last term of 
Eq.\eqref{eq:ZAV*-FFnew-2} and 
are proportional to the form factor combination $(2h_6^V \!-\!h_2^V)$.\ 
However, according to the ET identity \eqref{eq:ET}, the leading
energy dependence of  Eq.\eqref{eq:FT8-ZL-h6} should be
${O}(E^3)$ only.\ Hence there must be an {\it exact cancellation} between
the ${O}(E^5)$ terms associated with the form factors $h_2^V$ and $h_6^V$.\ 
This means that the last term of Eq.\eqref{eq:ZAV*-FFnew-2} must vanish,
leading to the condition:
\beq
\label{eq:h2=2h6}
h_2^V=\hs 2\hs h_6^V\hs.
\eeq 
Using this condition, we express the CPV nTGC vertex 
\eqref{eq:ZAV*-FFnew} in the following form:
\begin{align}
\label{eq:ZAV*-FFnew2}
\Gamma_{Z\gamma V^*}^{\alpha\beta\mu}
&=	\frac{~e\hs (q_3^2\hsm -\! M_V^2)\,}{\,M_Z^{2}~}\!
\!\[ h_1^V\hsmx \big(q_2^\alpha g^{\mu\beta}\hsm\!-q_2^\mu g^{\alpha\beta}\big)
\hsmx +\! \frac{h_2^V}{\,2M_Z^2\,}\hs 
\hsm \big(M_Z^2\hs q_2^{\alpha}g^{\mu\beta}\!-q_3^2q_2^\mu g^{\alpha\beta}\big)\!\]\!.
\end{align}

In the next step, we match the newly-proposed form factor formulation of the nTGC vertices 
\eqref{eq:ZAV*-FFnew2} 
with the CPV nTGC vertices \eqref{eq:ZAV-dim8WB}-\eqref{eq:ZAV-dim8G+-} 
generated by the gauge-invariant dimension-8 CPV nTGC operators 
\eqref{eq:dim8H}-\eqref{eq:OG+G-} after the spontaneous electroweak symmetry breaking.\ 
From these, we derive a nontrivial relation between $h_2^Z$ and $h_2^\ga\,$:
\beq
h_2^Z=(c_W^{}\hsm /\hsm s_W^{})\hs h_2^\ga \,,
\eeq 
where we denote $h_2^Z\!\equiv\! h_2^{}\hs$.\
For convenience, we denote the polarization vectors 
for $Z_T^{}$, $Z_L^{}$ and $\ga$ by 
$\epsilon_{T\alpha}^{}$, $\epsilon_{L\alpha}^{}$ and $\epsilon_{\beta}^{}$ respectively.\  
Because $q_2^\alpha\epsilon_{T\alpha}^{}\!=\!0\hs$,
the structure $q_2^\alpha g^{\mu\beta}$ has a nonzero contribution only for the external state $Z_L^{}$,
but vanishes for the external state $Z_T^{}$.\  
On the other hand, as explained below Eq.\eqref{eq:ZAV*-FFnew}, the contraction 
$g^{\alpha\beta}\epsilon_{L\alpha}^{}\epsilon_{\beta}^{}\!=\!0\hs$.\ 
Thus, the structure $q_2^{\mu} g^{\al\be}$ has a nonzero contribution from the external state
$Z_T^{}$ and not $Z_L^{}\hs$.\

The kinematics for the reaction
$\hs e^+\hs e^-\!\ito Z\hs\ga\!\ito\! f\bar f\ga\hs$
are defined by the three angles  
$(\theta,\,\theta_*^{},\,\phi_*^{})$,
where $\,\theta\,$ is the polar scattering angle between 
the direction of the outgoing $Z$ and 
the initial state $e^-$,
$\,\theta_*^{}$ denotes the angle between the
direction opposite to the final-state $\gamma$
and the final-state fermion $f$ direction in the $Z$ rest frame,
and $\,\phi_*^{}\,$ is the angle between the scattering plane
and the decay plane of $Z$ in the $\hs e^+\hs e^-\hs$ 
center-of-mass frame.

With these definitions, we derive the following contributions of the CPV nTGC form factor vertices \eqref{eq:ZAV*-FFnew2} 
to the scattering amplitudes for 
$\hs f_{\lam}^{}\bar f_{\lam'}^{}\ito Z\ga\hs$: 
\beqs
\label{eq:T8h}
\begin{align}
& \hspace*{-6mm}
\mathcal{T}_{(8),\text{F}}^{ss'\!,\text{T}}\!\!\left\lgroup\!\!\!
\begin{array}{cc}
-- \!&\! -+ \\
+- \!&\! ++\\
\end{array}\!\!\!\right\rgroup\!\!
=\frac{~\ii\hs e^2 \sin\theta(c^{\hs\prime V}_L\!+\! c^{\hs\prime V}_R)(M_Z^2\!-\!s)
\big(2 h_1^V\hsm M_Z^2\!+\!h_2^V\hsm s\big)\,}
{4 M_Z^4\hs s_W^{}c_W^{}}
\!\(\!\!
\begin{array}{cc}
	1 & 0 \\
	0 & 1 \\
\end{array} \!\!\) \!,
\\[1mm]
& \hspace*{-6mm}
\mathcal{T}_{(8),\text{F}}^{ss'\!,\text{L}}(0-,0+)
=-\frac{\,\ii\hs e^2 (2 h_1^V\!\!+\hsm h_2^V) s^{\frac{1}{2}} (s\!-\hsm\!M_Z^2)\,}
{2 \sqrt{2\,} M_Z^3\hs s_W^{}c_W^{}}\!\!
\(\hsm\! c^{\hs\prime V}_L \!\hsm\sin^2\!\hsm\frac{\,\theta\,}{2}
\!-\hsm c^{\hs\prime V}_R \!\hsm\cos^2\!\hsm\frac{\,\theta\,}{2},\, 
c^{\hs\prime V}_L\!\hsm\cos^2\!\hsm\frac{\,\theta\,}{2}
\!-\!c^{\hs\prime V}_R \!\hsm\sin^2\!\hsm\frac{\,\theta\,}{2}\!\hsm\)\!,
\end{align}
\eeqs
for the helicity combinations
$\lam\lam'\!=\!(--,-+,+-,++)$ and $\lam\lam'\!=\!(0-,0+)$,
where the subscript F for each amplitude $\TT$ denotes the contributions from form factors.\  
In the above, the coupling coefficients are defined as follows: 
\\[-6mm]
%
\begin{align}
c_L^{\hs\prime Z} & = c_L^{Z}\delta_{\!s,-\frac{1}{2}}^{} \hs, 
\hspace*{8mm}
c_R^{\hs\prime Z} = c_R^{Z}\hs\delta_{\!s,\frac{1}{2}} \hs, 
\hspace*{7mm}
c_{L,R}^{\hs\prime\hs\ga} = c_{L,R}^{\ga}\hs \delta_{\!s,\mp\frac{1}{2}} \,, 	
\nn\\
c_L^{Z} & = I_3^{}\hsm -\hsm Qs_W^2 \hs,
\hspace*{5.4mm}
c_R^{Z} = -Qs_W^2 \hs,
\hspace*{5.8mm}
c_{L,R}^{\ga} = Q\hs s_W^{}c_W^{} \,, 	
\end{align}
%
where we have used the notations   
$\,(s_W^{},\,c_W^{})\!=\!(\sin\hsm\theta_W^{},\hs \cos\hsm\theta_W^{})$.\  
In the above, the subscript index
$\,s \!=\!\mp\frac{1}{2}$\,
denotes the initial-state fermion helicities, 
$I_3^{}$ is the weak isospin, and $Q$ is the electric charge.\  

\vs

We recall that the SM contributions to the helicity amplitudes of 
$\hs f_{\lam}^{}\bar f_{\lam'}^{}\ito Z\ga\hs$ 
have been computed in the previous studies\,\cite{Ellis:2022zdw}\cite{Ellis:2020ljj}\cite{Ellis:2019zex}:
\beqs
\label{eq:Tsm-T+L}
\begin{eqnarray}
	\mathcal{T}_{\text{SM}}^{ss'\!,\text{T}}\!\!\left\lgroup\!\!\!
	\begin{array}{cc}
		-- \!&\! -+ \\
		+- \!&\! ++\\
	\end{array}\!\!\right\rgroup
	\hspace*{-2.5mm}
	\!\!&=&\!\!\!
	\frac{-2\hs e^2 Q}{\,s_W^{}c_W^{}(s\!-\!M_Z^2)\,}\!\!\!
	\left\lgroup\!\!\!\!
	\begin{array}{ll}
		\left(c_L'\!\cot\!\frac{\theta}{2}\!-\!
		c_R'\!\tan\!\frac{\theta}{2}\right)\!M_Z^2~
		\!&\!\!
		\left(-c_L'\!\cot\!\frac{\theta}{2}\!+\!
		c_R'\!\tan\!\frac{\theta}{2}\right)\!s
		\\[2mm]
		\left(c_L'\!\tan\!\frac{\theta}{2}\!-\!
		c_R'\!\cot\!\frac{\theta}{2}\right)\!s
		\!&\!\!
		\left(-c_L^{}\!\tan\!\frac{\theta}{2}\!+\!c_R^{}\!\cot\!\frac{\theta}{2}\right)\!M_Z^2
	\end{array}
	\!\!\!\right\rgroup\!\!,\hspace*{14mm}
	\label{msmT}
	\label{eq:Tsm-T}
\\[2mm]
	\mathcal{T}_{\text{SM}}^{ss'\!,\text{L}}(0-,0+)
	\hsm\!\!\!&=&\!\!\! \frac{\,-2\sqrt{2}\hs 
		e^2Q\hs (c_L'\!\!+\!c_R')\hs M_Z^{}\sqrt{s\,}~}
	{\,s_W^{}c_W^{}(s\!-\!M_Z^2)\,}\left(1,\,-1\right),
	\label{msm}
	\label{eq:Tsm-L}
\end{eqnarray}
\eeqs
where the coupling coefficients are given by
\beq 
c_L' = (I_3^{}\hsm -\hsm Qs_W^2)\delta_{\!s,-\frac{1}{2}}^{}\hs, \hspace*{5mm}
c_R' = -Qs_W^2\delta_{\!s,\frac{1}{2}}^{} \hs. 
\eeq 

In passing, we note that 
the conventional CPV nTGC form factors have been used to study possible 
probes of CPV nTGCs at $e^+e^-$ colliders 
in the literature\,\cite{Rahaman:2016pqj}\cite{Ananthanarayan:2014sea}.\ 
However, the analysis of Ref.\,\cite{Rahaman:2016pqj} was restricted
to dimension-6 operators in the broken electroweak symmetry phase,
which respects only the residual U(1)$_{\rm{EM}}^{}$ gauge symmetry
and is incompatible with the dimension-8 SMEFT formalism for
the nTGC operators, contrary to our new CPV nTGC form factor 
formulation \eqref{eq:ZAV*-FFnew} and 
our CPC nTGC form factor approach in Ref.\,\cite{Ellis:2022zdw}.\ 
Also, Ref.\,\cite{Ananthanarayan:2014sea} did not consider 
$Z$-decay final states, so the CPV nTGC part of the 
amplitude in their analysis is imaginary and
cannot interfere with the SM part
(whose CPV form factors are real).\
Ref.\,\cite{Ananthanarayan:2014sea} assumed these CPV form factors 
to be imaginary, which would give nonzero interference with the SM part.\ 
However, taking the CPV form factors to be imaginary is
incompatible with the Hermiticity of the Lagrangian.

\hspace*{0.5mm}
\section{\hspace*{-2.5mm}Analyzing Sensitivities to CPV nTGCs at {$\mathbf{e^+e^-}$} Colliders}
\label{sec:3} 
\label{sec:sensitivities}

In this Section we analyze the sensitivity reaches for
probes of the CPV nTGCs at future $e^+e^-$ colliders via the reaction 
$\hs e^+e^-\!\hsm\ito Z\gamma\hs$ with $Z\ito \ell\bar{\ell},q\bar{q}$ decays.\
As we discuss later, in the case of the invisible decay channel 
$Z\!\ito \nu\bar{\nu}$ 
the interference term in the cross section cannot be measured, 
which results in weaker sensitivities.

\begin{figure}[t]
\centering
\includegraphics[height=9cm,width=11cm]{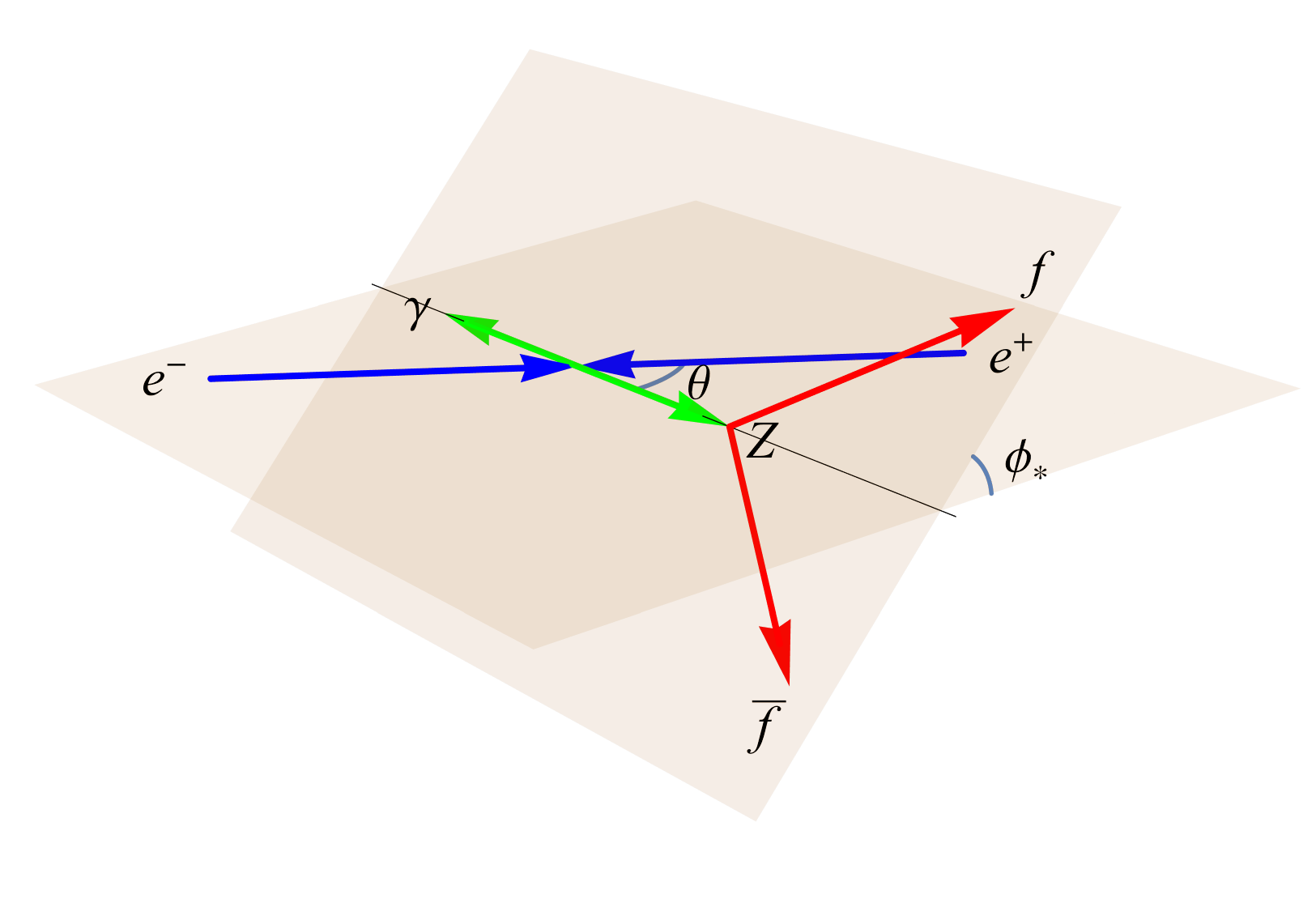}
\vspace*{-11mm}
\caption{\small 
Kinematics in the $e^+e^-$ collision frame of the reaction $e^+e^-\!\!\to\!Z\gamma$
followed by $Z\!\ito\!f\bar f$ decay.\
In this plot, $\theta$ denotes the polar scattering angle between the directions of the outgoing $Z$ 
and the initial state $e^-$, and $\phi_*$ is defined as the angle between the scattering plane 
and the decay plane of the final state $Z$ boson in the center-of-mass frame of $e^-e^+$. }
\label{fig:phi_angle}
\hspace*{1mm}
\end{figure}

\subsection{\hspace*{-4mm}
Cross Sections and Observables: Analysis of CPV nTGCs}
\label{sec:cross-sections}
\label{sec:3.1}

For the reaction $e^+e^-\!\!\to\!Z\gamma$ (with $Z\!\!\to\!\!f\bar f$ decays), 
we illustrate in Fig.\,\ref{fig:phi_angle} its kinematics in the $e^-e^+$ collision frame.\
In this plot, $\theta$ denotes the polar scattering angle between the directions of the outgoing $Z$ boson  
and the initial state electron $e^-$, whereas $\phi_*$ is defined as the angle between the scattering plane 
and the decay plane of the final state $Z$ boson in the center-of-mass frame of $e^-e^+$.\

The total cross section $\sigma$ for the on-shell process $e^+e^-\!\!\to\!Z\gamma$
can be divided into a SM part $\sigma_0^{}$, an interference term $\sigma_1^{}$, 
and a pure new phyics term $\sigma_2^{}\hs$.\ 
However, since the nTGC amplitudes with on-shell final state $Z\gamma$ 
are imaginary (having an extra factor of ``\,$\ii$\,''),  
there is no interference term in the cross section ($\hs\sigma_1^{}\!=\!0\hs$).\ 
Hence, we have the following expression for the total cross section $\sigma\,$:
\begin{align}
\label{eq:CS-qqZA-d8other}
\hspace*{-9mm}
\sigma(Z\gamma)
=&~ \sigma_0^{} + \sigma_2^{}
= \frac{~e^4(c_L^2\!+\!c_R^2)Q^2\!\hsm\left[-(s\!-\!M^2_Z)^2\!-\!2(s^2\!+\!M_Z^4)
	\ln\sin\!\frac{\,\delta\,}{2}\right]\,}
{\,8\hs\pi s_W^2c_W^2(s\!-\!M^2_Z)s^2\,}
\nn\\[1mm]
\hspace*{-9mm}
&
+\frac{\,\big[(c_L^V)^2\!+\!(c_R^V)^2\big]\!\(s\!-\!M^2_Z\)^{\!3}
\!\!\left\{\hsm \big[4(h_1^V)^2\hsm M_Z^2\!+\!(h_2^V)^2s\big]\!(s\!+\!M_Z^2)\!+\!8h_1^V\hsm h_2^V\!M_Z^2\hs s\right\}\,}
{\,768\hs\pi c_W^2s_W^2M_Z^8\,s^2\,} + O(\delta)\,,
\end{align}
\\[-14mm]
where $\,\delta\,$ is the lower cut on the scattering angle $\theta\hs$
and we have used the coupling notations 
$(c_L^{},\hs c_R^{})\!=\!(c_L^Z,\hs c_R^Z)\hs$.\ 
Then, we derive the expression of the scattering amplitude 
for the full process $e^+e^-\!\ito Z\ga\ito\hsm f\bar f \ga$
as follows:
\begin{align}
\label{eq:T-llgamma}
\hspace*{-8mm}
\mathcal{T}^{ss'}_{\si\si'\!\lambda}\hsm (f\bar f\ga) =\, &
\frac{~eM_Z^{}\mathcal{D}_Z^{}\,}{s_W^{}c_W^{}}
\!\!\left[\!\sqrt{2\,}e^{\ii\phi_*^{}}\!\!
\(\hsm\!f_R^{\si}\cos^2\!\frac{\,\theta_*^{}}{2}\!-\! f_L^{\si}\!\sin^2\!\frac{\,\theta_*^{}}{2}\!\)\!
\mathcal{T}_{ss'}^T(+\lam )
\right.
\\
\hspace*{-8mm}
& \left.
+\sqrt{2\,}e^{-i\phi_*^{}}\!\!
\(\hsm\!f_R^{\si}\sin^2\!\frac{\,\theta_*^{}}{2}\!
-\hsm f_L^{\si}\!\cos^2\!\frac{\,\theta_*^{}}{2}\!\)\!\hsm 
\mathcal{T}_{ss'}^T(-\lam )
\!+\! (f_R^{\si}\!+\!f_L^{\si})\sin\!\theta_* \mathcal{T}_{ss'}^L(0\lam )\hsm\right]\!,
\nn
\end{align}
where $\theta_*^{}$ denotes the angle between the direction opposite to the final-state $\gamma$ 
and the direction of the final-state fermion $f$ direction in the $Z$ rest frame.\ 
The coefficients
$f_L^\si \!=\! (I_3^{}\!-\hsm Qs_W^2)\delta_{\!\si,-\frac{1}{2}}^{}$ and 
$f_R^\si \!=\! (-Qs_W^2)\delta_{\!\si,\frac{1}{2}}$ 
above denote the couplings of the final-state fermions.\  
Using Eq.(\ref{eq:T-llgamma}), we can compute the interference term in 
the differential cross section and 
deduce its $\phi_*^{}$ dependence as follows:
\beq
\label{eq:ds1}
\frac{\d^3\sigma_1}{\,\d\theta\hs\d\theta_*\d\phi_*^{}\,}
\,=\, c_1^{}(\theta,\theta_*)\sin\phi_*^{} + c_2^{}(\theta,\theta_*)\sin2\phi_*^{} \, ,
\eeq
where the coefficients $c_1^{}(\theta,\theta_*)$ and $c_2^{}(\theta,\theta_*)$ 
are functions of $(\theta,\theta_*)$.

\vs 

We express each cross section term  
$\sigma_i^{}(e^+e^-\!\ito Z\ga \ito\hsm f\bar f\ga)\hsm\equiv\hsm  
\sigma_i^{}(e^+e^-\!\hsm\ito Z\ga)\text{Br}(f\bar f)$, 
and define the following angular distribution:
\begin{equation}
\tilde\sigma_i^{} \equiv 
\frac{~\d\sigma_i(e^+e^-\!\ito Z\ga\ito f\bar f\ga)~}
{\d\phi_*^{}\,\text{Br}(f\bar f)} \, ,
\end{equation}
where the contribution of the interference term is given by 
\begin{align}
\hspace*{-7mm}
\tilde\sigma_1^{}  
=~& \frac{~e^4 h_1^V Q \sin\phi_* (M_Z^2\!-\!s)~}{c_W^{}s_W^{}M_Z^2}\!\!
\left[\!\frac{~3(f_L^2\!-\!f_R^2) (M_Z^2\!+\!3 s)(c_L^V c_L^{Z}\!+\!c_R^V c_R^{Z})~}
{2048(f_L^2\!+\!f_R^2)s^{3/2}M_Z} - 
\frac{\cos\phi_*^{}(c_L^Vc_L^{Z}\hsm\!-\!c_R^Vc_R^{Z})~}{32\hs\pi^2s}
\!\right]
\nn
\\
\hspace*{-7mm}
& +\frac{~e^4 h_2^V Q \sin\phi_*^{}(s\!-\!M_Z^2)~}{c_W^{}s_W^{}M_Z^3}\!\!
\left[\!\frac{~3(f_L^2\!-\hsm\!f_R^2)(M_Z^2\!-\!5 s) 
(c_L^V c_L^{Z}\hsm\!+\!c_R^V c_R^{Z})~}{4096(f_L^2\!+\!f_R^2)\hs s^{3/2}}
+\frac{~\cos\phi_*^{} (c_L^V c_L^{Z}\hsm\!-\!c_R^V c_R^{Z})~}{64\hs\pi^2 M_Z}\hsm\right]
\nn\\
\hspace*{-7mm}
\simeq~& 
\frac{~9\sqrt{s\,}\hs e^4 h_1^V Q  (f_R^2\!-\!f_L^2) (c_L^V c_L^{Z}\hsm\!+\!c_R^V c_R^{Z}~)
\sin\phi_*^{}~}{2048\hs c_W^{}s_W^{}\hs M_Z^3 (f_L^2\!+\!f_R^2)~}
+\frac{~s\hs e^4 h_2^V Q (c_L^V c_L^{Z}\hsm\!-\!c_R^V c_R^{Z})\sin 2\phi_*^{}~}
{128\hs\pi^2c_W^{}s_W^{}M_Z^4},
\label{lastrow}
\end{align}
and the last row of Eq.\eqref{lastrow} gives the leading contributions 
in the limit $s\!\gg\! M_Z^2\hs$.
Since $\sigma_1^{}\!=\!\int\hsm\!\d\phi_*^{}\,
\tilde \sigma_1^{}\!=0\hs$, 
it is not convenient to define a normalized angular distribution 
for the interference term as was done for the analysis of the CPC nTGCs.

We may further define the normalized angular distribution functions as follows:
\begin{eqnarray}
	f_\xi^j \, = \, \frac{1}{\sigma_{\!j}^{}}\frac{\di\sigma_{\!j}^{}}{\,\di\xi\,}\,,
\end{eqnarray}
where the angles $\,\xi \in (\theta,\,\theta_*^{},\,\phi_*^{})$,\,
and the cross sections
$\,\sigma_j^{}$ ($j=0,1,2$) represent the SM contribution ($\sigma_0^{}$),
the ${O}(\cut^{-4})$ interference contribution ($\sigma_1^{}$), and the
${O}(\cut^{-8})$ contribution ($\sigma_2^{}$),\, respectively. 
For instance, we derive explicit formulae for the
normalized azimuthal angular distribution functions
$\,f_{\phi_*}^{0}\,$ and $\,f_{\phi_*}^{2}\,$ as follows:\footnote{%
Here we note that since $\sigma_1^{}$ vanishes, $f_\xi^{1}$ is not well defined.}
{\small
\beqs
\label{eq:f-phi-OGP}
\beqa
\label{eq:f0-phi}
\hspace*{-15mm}
f_{\phi_*^{}}^{0} \hspace*{-3mm} &=& \hspace*{-3mm} 
\frac{1}{\,2\pi\,}\hsm  +\hsm\frac{\,3\pi^2(c_L^2\!-\!c_R^2)(f_L^2\!-\!f_R^2)M_Z^{}\sqrt{s}\,
(s\!+\!M_Z^2)\cos\hsm\phi_*^{}\!
-8(c_L^2\!+\!c_R^2)(f_L^2\!+\!f_R^2)M_Z^2\,s \cos\hsm 2\phi_*^{}\,}
{\,16\hs\pi(c_L^2\!+\!c_R^2)(f_L^2\!+\!f_R^2)\!
			\left[(s\!-\!M_Z^2)^2\!+2(s^2\!+\!M_Z^4)\ln\sin\!\frac{\delta}{2}
			\hs\right]\,}+O(\delta) ,\hspace*{2mm}
\\[2mm]
\label{eq:f2-d-phi}
\hspace*{-15mm}
f_{\phi_*^{}}^{2} \hspace*{-3mm} &=& \hspace*{-3mm} 
\frac{1}{\,2\pi\,} 
-\frac{9 \hs\pi  \sqrt{s} M_Z \big(f_L^2\!-\!f_R^2\big) \big(c_L^V{}^2\hsm\!-\!c_R^V{}^2\big) 
(2 h_1^V\hsm\!+\!h_2^V)\big(s h_2^V\hsm\!+\!2 M_Z^2 h_1^V\big)\! \cos\phi_*{}}
{~128 \big(f_L^2\!+\!f_R^2\big) \big(c_L^V{}^2\!+\!c_R^V{}^2\big) 
\big[4 M_Z^2 h_1^V{}^2 (s\!+\hsm\!M_Z^2)\!+8\hs s\hsm M_Z{}^2 h_1^V h_2^V+s h_2^V{}^2 \big(s \!+\! M_Z^2\big)\big])}\, ,
\label{eq:fphi*-2}
\eeqa 
\eeqs  
}
\hspace*{-2.4mm}
where we denote the $Z$ couplings with the initial state quarks
as $\,(c_L^{},\,c_R^{}) \!=\! (I_3^{}\hsm -\hsm Qs_W^2,\hs -Qs_W^2)$,
and the $Z$ couplings with the final-state fermions as
$\hs (f_L^{}, f_R^{}) \hsm =\hsm
(I_3^{}\! -\! Qs_W^2,\hs -Qs_W^2)\hs$.\ 
In the above, $\delta$ is the lower cut on the scattering angle $\theta\hs$.\ 
We note that both the angular distributions $f_{\phi_*^{}}^{0}$ and $f_{\phi_*^{}}^{2}$ 
approach $\frac{1}{\,2\hs\pi\,}$
when $s\!\gg\! M_Z^2\,$.

\begin{figure}[t]
\centering
\vspace*{-6mm}
\includegraphics[height=5cm]{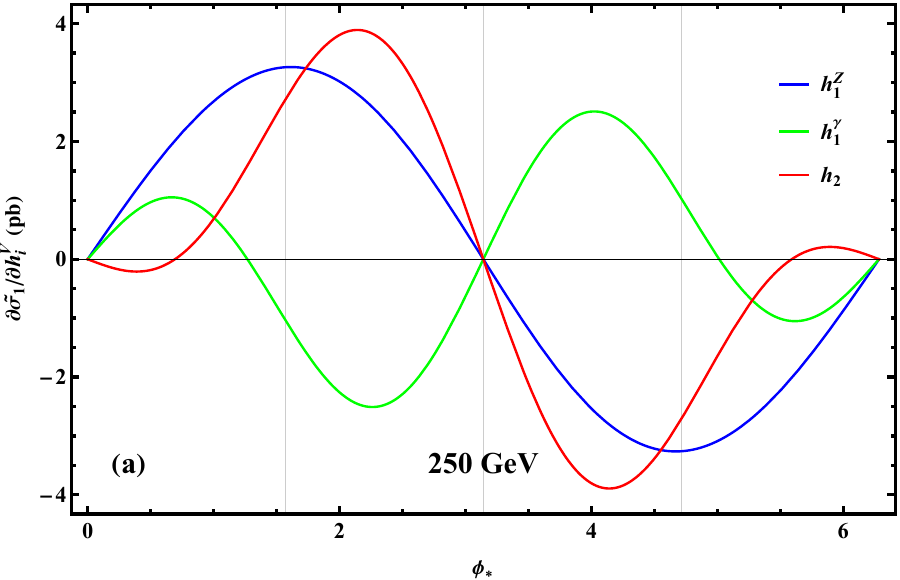}
\includegraphics[height=5cm]{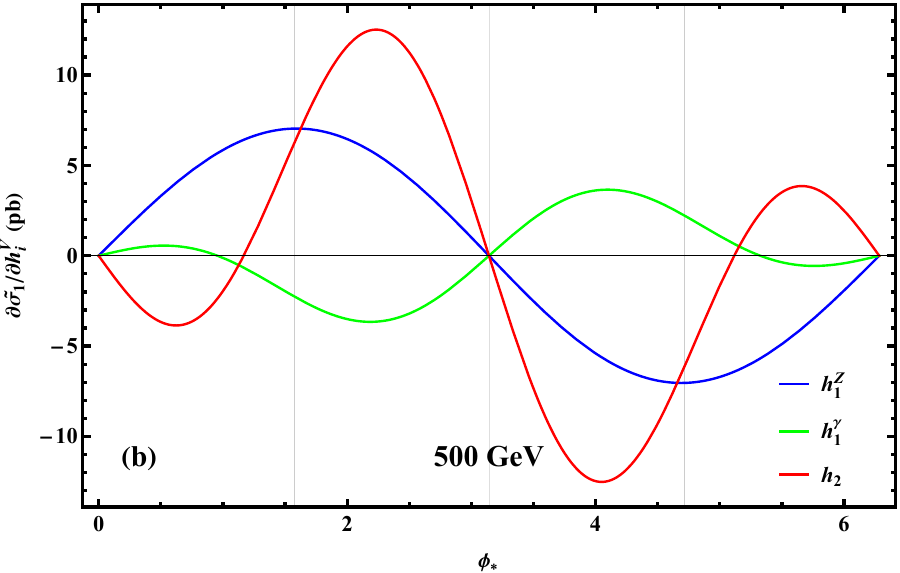}
\\
\includegraphics[height=5cm]{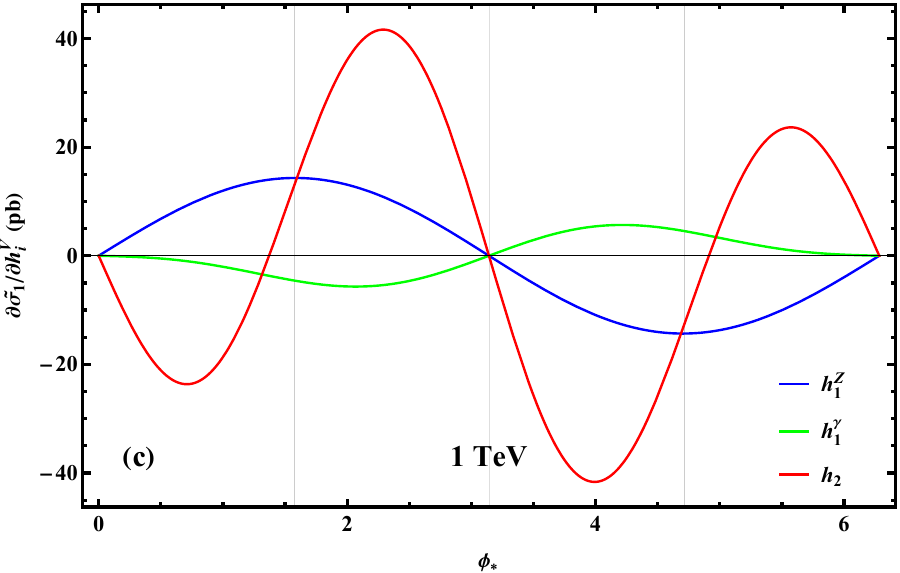}
\includegraphics[height=5cm]{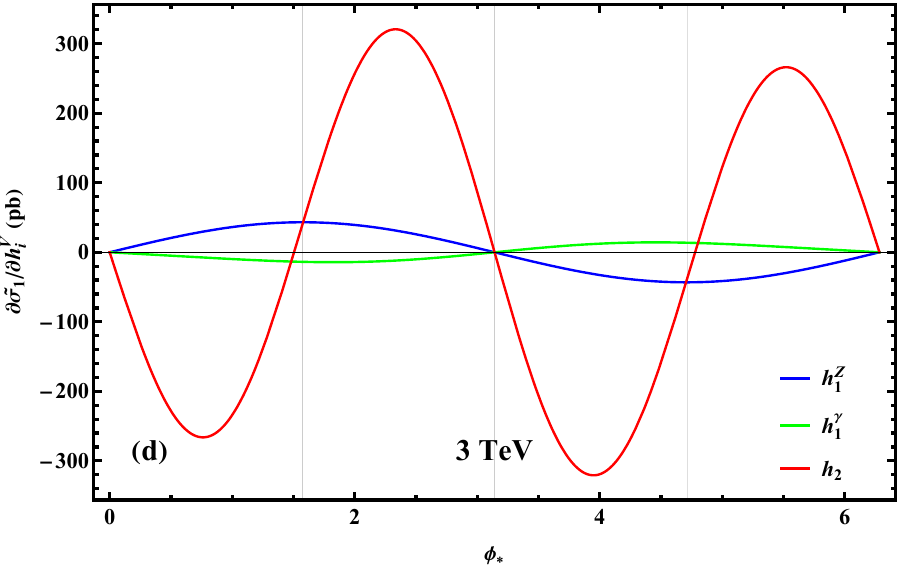}
\vspace*{-2mm}
\caption{\hspace*{0mm}\small 
Angular distributions in $\phi_*^{}$ 
for the reaction $\hs e^+\hs e^{-}\!\ito Z\ga\hs$ with $Z\!\ito d\hs\bar{d}\hs$,
as generated by the form factors $(h_1^Z,\, h_1^\ga,\, h_2^{})$ 
at $e^+e^-$ colliders.\  The plots\,(a)-(d) present the angular distributions for $e^+e^-$ collider energies
$\sqrt{s\,}\hsm =\hsm (0.25,\hs 0.5,\hs 1,\hs 3)\hs$TeV, respectively.
}
\label{fig:1}
\label{fig:2n}
\end{figure}
\noindent

In Fig.\,\ref{fig:1}, we present the angular distributions 
at different $e^+ e^-$ colliders 
with collision energies $\sqrt{s} = (0.25,\,0.5,\, 1,\,3)$\,{TeV}, respectively.\footnote{%
We note that our results apply equally to $\mu^+ \mu^-$ colliders.}\ 
The $h_1^V$ term with leading energy dependence is proportional to $\sin\phi_*^{}\hs$, 
whereas for $h_2^V$ it is proportional to  $\sin2\phi_*^{}\hs$.\  
We note that for the form factor $h_1^Z$, the coupling coefficient
$(c_L^Z)^2\hsm -\hsm (c_R^Z)^2\!=\!I_3^{}(I_3^{}\hsm - 2Qs_W^2)$ 
is suppressed by the factor
$(\frac{1}{2}\hsm -\hsm 2s_W^2)\simeq 4\%$ \cite{PDG}.\ 
Hence the $h_1^Z$ angular distribution is 
dominated by the contribution $\propto\sin\phi_*^{}\hs$.\  
On the other hand, the distributions for $h_1^\ga$ and $h_2^{}$ are combinations of 
$\sin\phi_*^{}$ and $\sin2\phi_*^{}$.\  
In the high-energy limit, the $h_1^\ga$ ($h_2^{}$) distribution is proportional to
$\sin\phi_*^{}$ ($\hs\sin2\phi_*^{}$), 
as shown in the last row of Eq.\eqref{lastrow} and in Fig.\,\ref{fig:2n}.\ 

\vs 

The sign of the interference term (\ref{eq:ds1}) depends on 
the angles $(\theta,\, \theta_*^{},\hs \phi_*^{})$.\ 
In order to avoid large cancellations between the contributions 
having different signs, we divide the phase space into different regions 
and compute the interference term for each region independently.\ 
To this end, we separate the ($\theta,\hs \theta_*^{},\hs \phi_*^{}$) 
phase space into $2\!\times\! 4\!=\!8$ regions 
with the boundaries $\cos\theta\cos\theta_*^{}\!=\hsm 0\hs$ and 
$\phi\!=\!\frac{\pi}{2},\hs\pi,\hs \frac{3\pi}{2}$,  
and denote the cross section $\sigma_i^{}$ in region $j$ as $\sigma_{i,j}^{}$.\  
Since $\dis\sum_{j=1}^{8}\!\sigma_{1,j}^{}\!=\!0\hs$, 
we analyse the signals in each region separately and combine their significances as follows:
\beqs
\begin{align}
\mathcal Z_f^{} &= 
\sqrt{\!\sum_j\!\frac{S_j^2}{ B_j}~}
=\sqrt{\!\sum_j\!\frac{\sigma_{1,j}^2}{ \si_{0,j}}\!\times\! \mL\!\times\!\epsilon~} \,,
\\
\mathcal Z &=
\sqrt{\sum_f\!\hsm Z_f^2~}\simeq \sqrt{3\mathcal Z_d^2\!+\!2\mathcal Z_u^2\!+\!3\mathcal Z_{\ell}^2~} \,,
\end{align}
\label{eq:z}
\eeqs
\hspace{-2.3mm}
where $\mL$ is the integrated luminosity, $\epsilon$ is the detection efficiency and $f$ denotes the type of fermion.\ 
In the high energy limit $s\!\gg\! M_Z^2\hs$, the sensitivity bound scales as  
$h_1^V\!\!\propto\! E^{-2}$ for probing $h_1^V$ 
and scale as $h_2^{}\!\propto\! E^{-3}$ for probing $h_2^V$, 
where we denote the center-of-mass energy $E\!=\!\!\sqrt{s\,}$.\ 
Accordingly, the sensitivity bounds for probing the new physics scales $\cut$ 
of dimension-8 operators are $\cut\!\propto\! E^{3/4}$ for $O_{\!\bar G+}^{}$ 
and $\cut\!\propto\! E^{1/2}$ for the other operators. 
This is why the sensitivity bound on probing $O_{\!\bar G+}^{}$ is more sensitive and it is enhanced  
faster with the increase of collision energy.\
Besides, since the significances $\mathcal Z_{\ell}^{}$ of the lepton channels contribute only around 5\% 
of the overall significance $\mathcal Z$, 
their contributions to the sensitivity reaches on probing the scale of new physics are negligible.\

In the above we have considered the irreducible SM backgrounds of $Z\gamma$ production 
from $t$-channel and $u$-channel fermion exchanges, having the on-shell decays of 
$Z\hsm\ito\hsm f\bar{f}$.\ 
We note that in principle 
the scattering process $e^+e^-\!\ito\hsm f\bar{f}\ga\,$ also contains other contributions 
whose final state fermions $f\bar{f}$ do not arise from the on-shell $Z$-decays, and thus serve as
the reducible SM backgrounds.\ 
These backgrounds are the same as what we studied in the analysis of CPC nTGCs\,\cite{Ellis:2019zex}\cite{Ellis:2020ljj}.\ 
By imposing the invariant-mass cut $|M_{f\bar f}^{} -\hsm M_Z^{}|\!<\!10\hs$GeV, 
the reducible backgrounds can be reduced to less than 10\% of the total backgrounds.\ 
Most $Z\gamma$ backgrounds can be removed by imposing lower cut on scattering angle $\theta$.\ 
The irreducible backgrounds are shown in Table\,\ref{tab:1new}, 
where we have imposed a lower cut on the scattering angle  $\theta\!>\!\delta$ (with $\delta\!=\!0.33$),
and an invariant-mass cut 
$|M_{f\bar f}^{} -\hsm M_Z^{}|\!<\!10\,$GeV.\  
Since the signal significance is $\mathcal{Z}\!=\!S/\!\sqrt{B\,}$, the sensitivity bounds on the nTGC
form factors $h_i^V$ scale as $h_i^V\!\!\propto\!\!\sqrt{B\,}$ 
and on the nTGC cutoff $\Lambda$ behave as $\Lambda\!\propto\! B^{-1/8}$.\ 
Thus the sensitivity bounds are insensitive to a change of backgrounds $B$ by less than 10\%.\ 
Namely, varying $B$ by less than 10\% only causes a change of the sensitivity bounds on $h_i^V$ by less than
5\% and on $\Lambda$ by less than 1.3\%.\ Hence it is justified to use the irreducible backgrounds 
for estimating sensitivity bounds.\

\begin{table}[b]
\centering
\begin{tabular}{c||c|c|c|c|c}
\hline\hline 
		&&&&&
		\\[-3.5mm]
		$\sqrt{s\,}$ (TeV) & 0.25 & 0.5 & 1 & 3 & 5
		\\
		\hline\hline 
		&&&&&
		\\[-4mm]
		$Z\!\to\!d_j^{}\bar{d}_j^{}$ & 780 & 160 & 38.3 & 4.20 & 1.51
		\\[0.7mm]
		\hline
		&&&&&
		\\[-4.3mm]
		$Z\!\to\!u_j^{}\bar{u}_j^{}$ & 604 & 124 & 29.8 & 3.26 & 1.17\\
		\hline
		&&&&&
		\\[-4mm]
		$Z\!\to\!\ell_j^{}\bar{\ell}_j^{}$ & 176 & 36.5 & 8.70 & 0.96 & 0.34\\
		\hline\hline 
\end{tabular}
\vspace*{-1mm}
\caption{\small Cross section  (in fb) of the SM irreducible backgrounds
for the reaction $e^+e^-\!\!\hsm\to\hsm\!Z\gamma$ with different $Z$ decay channels 
$Z\!\hsm\to\! d_j^{}\bar{d}_j^{},u_j^{}\bar{u}_j^{},\ell_j^{}\bar{\ell}_j^{}$,
having the final state down-quarks $d_j^{}\!\!=\!d,s,b$, up-quarks $u_j^{}\!\!=\hsm\! u,c$,
and leptons $\ell_j^{}\!=\!e,\mu,\tau$.\   
}
\label{tab:bkg}
\label{tab:1new}
\end{table}

\begin{table}[]
\centering
{\small
			\begin{tabular}{c||c|c|c}
				\hline\hline
				&&\\[-3.5mm]				
				$\sqrt{s\,}$\,(TeV)\, 
				& $|h_2^{}|$(unpol,{\hs}pol) & $|h_1^Z|$(unpol,{\hs}pol) & $|h_1^\gamma|$(unpol,{\hs}pol)
				\\
				&& 
				\\[-3.7mm]
				\hline\hline
				&& 
				\\[-3.8mm]
				0.25   & (2.7,1.4)$\times10^{-4}$
				& (3.1,\hs 2.4)$\times10^{-4}$
				& (3.9,\hs 1.6)$\times10^{-4}$					
				\\
				&& \\[-3.8mm]
				\hline
				&& \\[-3.8mm]
				0.5 &	(3.6,\hs 1.8)$\times10^{-5}$ 
				& (6.5,\hs 5.1)$\times 10^{-5}$ & (8.8,\hs 3.4)$\times10^{-5}$  
				\\
				&& \\[-3.8mm]
				\hline
				&& \\[-3.8mm]
				1 &	(4.7,\hs 2.4)$\times10^{-6}$ 
				& (1.6,\hs 1.2)$\times10^{-5}$ & (2.1,\hs 0.82)$\times10^{-5}$
				\\
				&& \\[-3.8mm]
				\hline
				&& \\[-3.8mm]
				3 & (1.7,\hs 0.87)$\times10^{-7}$
				&(1.7,\hs 1.3)$\times10^{-6}$ & (2.4,\hs 0.90)$\times10^{-6}$ 
				\\
				&&\\[-3.8mm]
				\hline
				&& \\[-3.8mm]
				5 &\,	(3.7,\hs 1.9)$\times10^{-8}$ 
				\,&\, (6.2,\hs 4.8)$\times10^{-7}$ 
				\,&\, (8.6,\hs 3.3)$\times10^{-7}$ \,
				\\
\hline\hline
\end{tabular}
\vspace*{-1mm}
\caption{\small\hspace*{-1mm}
Sensitivity reaches ($\hs 2\hs\sigma\hs$) for the CPV nTGC form factors from measuring the reaction
$\,e^-e^+\!\!\to\!Z\ga$ (with $Z\!\!\to\! q\bar{q}\hs ,\hs \ell\bar{\ell}\hs)$
				for \hs$e^+e^-\!\!$ colliders with different collision energies, 
				assuming $\mL\!=\!5$\,ab$^{-1}$ {in each case,
				for unpolarized $e^\mp$ beams and polarized
				$e^\mp$ beams with $(P_L^e,\hs P_R^{\bar e})\!=\!(0.9,\hs 0.65)$.\ 
				We impose a lower cut on the scattering angle $\theta\!>\!\delta$ (with $\delta\hsm =\hsm 0.33$)  
				and the near-on-shell condition 
				$|M_{f\bar f}^{}\!-\!M_Z^{}|\!<\!10$}\,GeV.}
\vspace*{1.5mm}
\label{tab:2new}
\label{tab:1}
\label{tab:f-p}}  	
\end{table}

\begin{table}[]
{\small
\centering
\begin{tabular}{c||c|c|c|c|c}
				\hline\hline
				&&&&&\\[-3.5mm]				
				$\sqrt{s\,}$\,(TeV)\, 
				& $\cut_{\bar G+}$(unpol,{\hs}pol)& $\cut_{\bar G-}$(unpol,{\hs}pol) & $\cut_{WW}$(unpol,{\hs}pol) & $\cut_{WB}$(unpol,{\hs}pol) & $\cut_{BB}$(unpol,{\hs}pol)
				\\
				&&&&& \\[-3.7mm]
				\hline
				&&&&& \\[-3.8mm]
				0.25   & (1.2,\hs 1.4) & (0.80,\hs 1.0) & (0.83,\hs 0.98 )	& (0.98,\hs 1.2) & (1.3,\hs 1.4)				
				\\
				&&&&& \\[-3.8mm]
				\hline
				&&&&& \\[-3.8mm]
				0.5  &	(2.0,\hs 2.4) & (1.2,\hs 1.5) & (1.2,\hs 1.5) & (1.4,\hs 1.8) & (1.8,\hs 2.0)			
				\\
				&&&&& \\[-3.8mm]
				\hline
				&&&&& \\[-3.8mm]
				1  & (3.3,\hs 4.0)	&(1.6,\hs 2.1) & (1.8,\hs 2.1) & (2.1,\hs 2.6) & (2.6,\hs 2.9)
				\\
				&&&&& \\[-3.8mm]
				\hline 
				&&&&& \\[-3.8mm]
				3  &(7.7,\hs 9.1) & (2.9,\hs 3.7) & (3.1,\hs 3.6) & (3.6,\hs 4.5) & (4.5,\hs 5.0)
				\\
				&&&&&\\[-3.8mm]
				\hline
				&&&&& \\[-3.8mm]
				5  & (11,\hs 13) & (3.7,\hs 4.7) & (3.9,\hs 4.7) & (4.6,\hs 5.8) & (5.8,\hs 6.5)
				\\
				\hline\hline
\end{tabular}
\vspace*{-0.5mm}
\caption{\small\hspace*{-1.5mm}
{Sensitivity reaches} ($\hs 2\hs\sigma\hs$) {for the new physics scale of the CPV nTGC operators  
from measurements of the reaction 
$\,e^-e^+\!\!\hsm\to\hsm\!Z\ga$ (with $Z\!\hsm\to\! q\bar{q}\hs ,\hs \ell\bar{\ell}\hs)$
for \hs$e^+e^-\!$ colliders with different collision energies, assuming} $\mL\!=5$\,ab$^{-1}$ 
{in each case, for the unpolarized $e^\mp$ beams and polarized $e^\mp$ beams with 
$(P_L^e,\hs P_R^{\bar e})\!=\!(0.9,\hs 0.65)$.\  
We impose a lower cut on the scattering angle $\theta\!>\!\delta$ (with $\delta\hsm =\hsm 0.33$)  
and the near-on-shell condition $|M_{f\bar f}^{}\!-\!M_Z^{}|\!<\!10$}\,GeV.
}
\label{tab:3new}
\label{tab:2}
\label{tab:o-p}	}
\end{table}

\begin{figure}[]
\vspace*{-5mm}
\centering
\includegraphics[height=8cm]{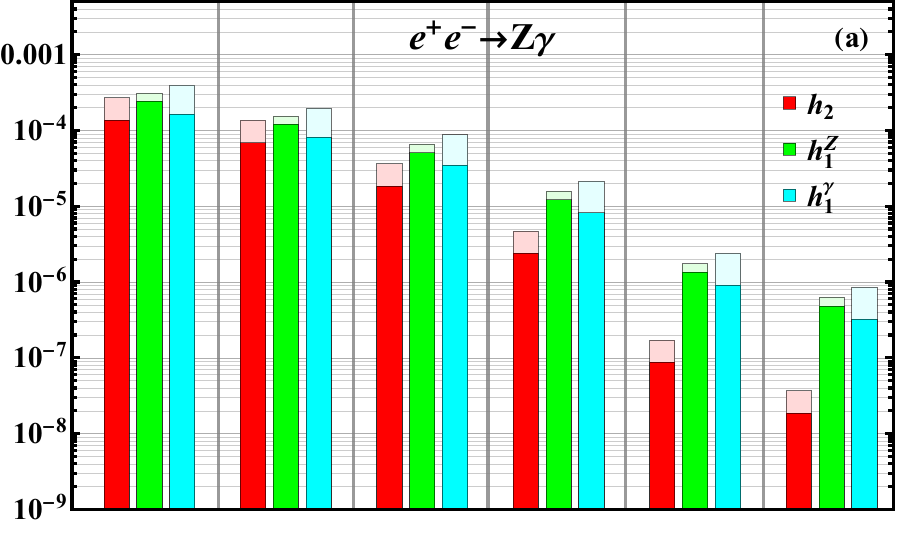}
\\
\includegraphics[height=8.8cm]{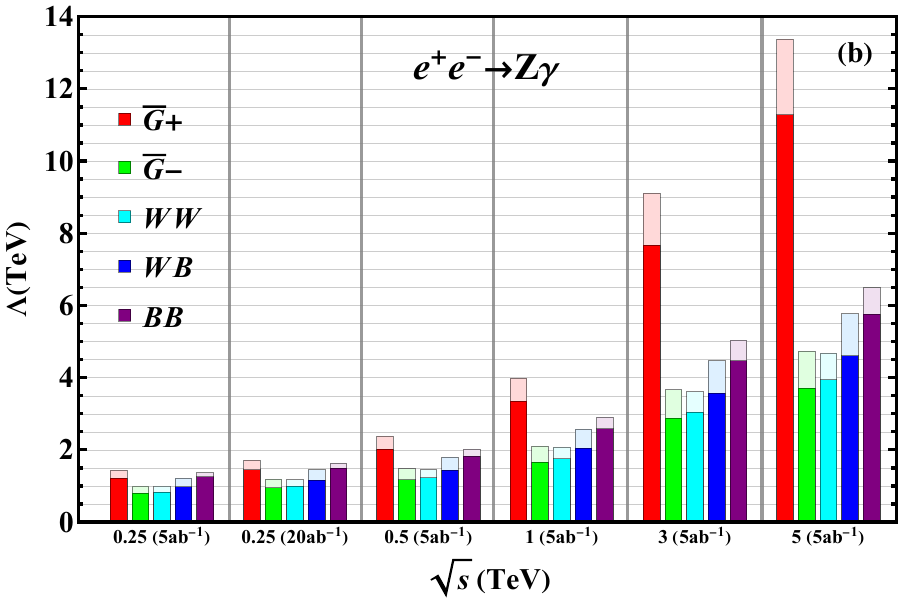}
\vspace*{-1mm}
\caption{\small\hspace*{-1mm}
{Sensitivity reaches ($\hs 2\hs\sigma$ bounds) on the CPV nTGV form factors 
			$(h_2^{},\hs h_1^Z,\hs h_1^\gamma)$ in plot\,(a) 
			and on the new physics scales $\cut$ of the corresponding dimension-8 
			CPV nTGC operators in plot\,(b) 
			at $e^+e^-$\! colliders with collision energies} 
		$\sqrt{s\,}\!=\!(0.25,\,0.5,\,1,\,3,\,5)$\,TeV, 
{by choosing integrated luminosities of $5\hs$}ab$^{-1}$ 
{in each case.\ For the case of $\sqrt{s\,}\!=\!250$\,GeV, we also present the sensitivity reaches with
an integrated luminosity of} $20\hs$ab$^{-1}$.\ 
{The sensitivities with unpolarized (polarized) beams 
are shown in light (heavy) colors in plot\,(a) and heavy (light) colors 
in plot\,(b) respectively, where we choose $(P_L^e,\hs P_R^{\bar e})\hsm =(0.9,\hs 0.65)$.}
}
\label{fig:bar}
\label{fig:3new}
\vspace*{1mm}
\end{figure}

\vs  
 
Fig.\,\ref{fig:bar} displays as histograms the 2$\sigma$ sensitivity bounds on the CPV 
nTGC form factors $(h_2^{},\hs h_1^Z,\hs h_1^\gamma)$ (upper panel)
and on the cutoff scales $\cut$ (lower panel) of the dimension-8 CPV nTGC operators   
at $e^+e^-$\! (or $\mu^+\mu^-$\!) colliders with collision energies 
$\sqrt{s\,}\!=\!(0.25,\,0.5,\,1,\,3,\,5)\hs$TeV, by choosing an integrated luminosities of 5\,ab$^{-1}$
in each case.\   
For the case of $\sqrt{s\,}\!=\!250$\,GeV, we also present the sensitivity reaches with
an integrated luminosity of $20\hs$ab$^{-1}$ \cite{CEPCStudyGroup:2023quu}.\ 
The sensitivities found for unpolarized (polarized) beams 
are shown in light (heavy) colors in the upper panel and
in  heavy (light) colors in the lower panel, respectively.
We see in the upper panel that the sensitivities to the form factors $h_1^{Z, \ga}$
recommended in our dimension-8 SMEFT approach are significantly weaker than the
sensitivities to the deprecated conventional form factor $h_2$ as the center-of-mass
energy increases. This feature can be traced back to the unphysical high-energy behavior
of the conventional parametrization. In the lower panel of Fig.~\ref{fig:bar} we see
that the sensitivity to the scale parameter $\cut$ for the dimension-8 operator $\tilde \mO_{G+}$
is much greater than those to the scale parameters for the
other operators $\tilde \mO_{G-}, \tilde \mO_{BW}, \tilde \mO_{WW}$
and $\tilde \mO_{BB}$.

\vs 

We present the sensitivity reaches for  
the different form factors in Table\,\ref{tab:f-p}, 
and on the operator scales in Table\,\ref{tab:o-p}.
In the case of $\mO_{\!\bar G^{}+}^{}$, 
we see that the operator scale sensitivity is more than 
1\,TeV at $\sqrt{s} \!=\! 250$\,GeV
and greater than 10\,TeV at $\sqrt{s\,} \!=\! 5$\,TeV, 
whereas the sensitivity scales for probing new physics in the
other dimension-8 operators range from several hundred GeV at $\sqrt{s\,} \!=\! 250$\,GeV
to several TeV at $\sqrt{s} \!=\! 5$\,TeV.\ 
The sensitivity reaches have the same order of magnitude as 
we found for the CPC case in Refs.\,\cite{Ellis:2022zdw,Ellis:2020ljj}.\ 
However, we note that the sensitivity
to $h_1^\ga$ is weaker than that to $h_1^Z$.\ 
This is because the total contributions of 
left-handed and right-handed electrons are proportional to the coupling 
$c_L^{}c_L^V\!+\hsm c_R^{}c_R^V$, 
and we note 
$|c_L^{}c_L^\ga \hsm + c_R^{}c_R^\ga|\!\ll\!|c_L^{}c_L^Z\hsm +\hsm c_R^{}c_R^Z|$
which is $0.0092\!\ll\! 0.13$ numerically.\ 
Since $\mO_{\bar G-}$ only contributes to $h_2^\ga$,
it has the weakest probing sensitivity among all the CPV nTGC operators.

We compare in Table\,\ref{tab:c} the sensitivities for probing $h_2^{}$ and 
the conventional form factors $h_1^Z$ and $h_1^\ga$.
The sensitivities to the conventional
form factors $h_2^Z$ and $h_1^\ga$ (marked in blue) 
are generally stronger than that for the SMEFT form factor
$h_2^{}$ (marked in red) by large factors, ranging
from ${O}(2)$ at $\sqrt{s}\!=\!250$\,GeV to ${O}(40)$ at a 
$\sqrt{s}\!=\!5$\,TeV $e^+e^-$ collider.\  
We note also that the sensitivity to $h_2^\ga$ is weaker than that to $h_2^Z$ 
for the same reason that the sensitivity to $h_1^\ga$ is weaker than that to $h_1^Z$. 
We emphasize again that the use of the conventional form factors is strongly deprecated,
because they are not compatible with the full electroweak gauge group SU(2)$_\rm{L}^{}\otimes$U(1)$_\rm{Y}^{}$ 
with spontaneous symmetry breaking and their apparent higher sensitivity is an artefact:
the blue numbers are included here only for comparison.

Using polarized $e^+e^-$ beams can improve the probing sensitivities.\  
We use the symbols $P_L^e$ ($P_R^{\bar e}$) to denote the fraction of 
the left-handed electrons (right-handed positrons) 
in the $e^-$\,($e^+$) beam.\footnote{%
We note that the degree of longitudinal beam polarization for electron $e^-$ or positron $e^+$
is defined as $\,\widehat{P}\!=\!P_R^{}\!-\!P_L^{}$ \cite{eePol}.\ 
Since the sum of left-handed and right-handed fractions equals one, $P_L^{}\!+\!P_R^{}\!=\!1\hs$,
the left-handed and right-handed fractions of $e^-\!$ or $e^+$ can be derived as follows,
$P_{L,R}^e\!=\!\fr{1}{2}(1\!\mp\!\widehat{P}^e)$\, and
$P_{L,R}^{\bar{e}}\!=\!\fr{1}{2}(1\!\mp\!\widehat{P}^{\bar{e}})$.\ 
For instance, an unpolarized beam of $e^-$ ($e^+$) has
a vanishing degree of polarization $\,\widehat{P}^e\!=\!0$ ($\hs\widehat{P}^{\bar{e}}\!=\!0\hs$)\,,
	whereas a polarized $e^-$ beam with a fraction $P_L^e=90\%$ has $\,\widehat{P}^e\!=\!-0.8$\,
	and a polarized $e^+$ beam with a fraction $P_R^{\bar{e}}=65\%$
	has $\,\widehat{P}^{\bar{e}}\!=\!0.3$\,.}\
The cross sections in the polarized case 
can be obtained from the unpolarized case by rescaling 
$(c_L^{},\hs c_L^V)\hsm\ito 2\sqrt{P_L^eP_R^{\bar e}\,}(c_L^{},c_L^V)$ and 
$(c_R^{},\hs c_R^V)\hsm\ito 2\sqrt{(1\!-\!P_L^e)(1\!-\!P_R^{\bar e})\,}(c_R^{},\hs c_R^V)\hs$.\  
We present the sensitivity bounds on the CPV nTGC form factors in Table\,\ref{tab:2new}
and on the cutoff scales of the CPV nTGC operators of dimension-8 in Table\,\ref{tab:3new}, 
for a sample input of $(P_L^e,\hs P_R^{\bar e})=(0.9,\hs 0.65)$.\  
We see that the sensitivities are improved 
by factors about 2 for probing the form factor $h_2^{}$ 
and are improved by about $30\%$ for probing $h_1^Z$, 
whereas those for probing $h_1^\ga$ are improved by larger factors about $2.6\hs$.\


\begin{table}[]
\centering
\begin{tabular}{c||c|c|c|c|c}
\hline\hline
			&&&&&\\[-3.5mm]				
			$\sqrt{s\,}$\,(TeV)\, 
			& 0.25&0.5&1&3&5
			\\
			&&&&& \\[-3.7mm]
			\hline
			&&&&& \\[-3.8mm]
			$|h_2|$&\red 2.7$\times10^{-4}$&\red	3.6$\times10^{-5}$ &\red	4.7$\times10^{-6}$ &\red 1.7$\times10^{-7}$&\red	3.7$\times10^{-8}$

			\\
			&&&&& \\[-3.8mm]
			\hline
			&&&&& \\[-3.8mm]
			$|h_2^Z|$	& \blu 1.1$\times10^{-4}$ & \blu 4.6$\times10^{-6}$ &\blu 2.7$\times10^{-7}$  &\blu 3.2$\times10^{-9}$ &\blu 4.2$\times10^{-10}$ 
			\\
			&&&&& \\[-3.8mm]
			\hline
			&&&&& \\[-3.8mm]
			$|h_2^\ga|$ &\blu  1.4$\times10^{-4}$	 
			&\blu  6.3$\times10^{-6}$&\blu  3.7$\times10^{-7}$&\blu4.4$\times10^{-9}$&\blu5.7$\times10^{-10}$
			\\
			\hline\hline
\end{tabular}
\vspace*{-1mm}
\caption{\small%
{Comparisons of the sensitivity reaches} ($2\hs\sigma$)
for the form factor $\hs h^{}_2\hs$ formulated in 
the SMEFT (marked in red color) and the deprecated conventional form factors $h_2^V\!$ 
respecting only U(1)$_{\rm{EM}}^{}$ (marked in blue color),
derived from analyses of the reaction
$\,e^+e^-\!\ito\hsm Z\ga$ (with $Z\ito \ell\bar{\ell},q\bar{q}$) at		
\hs$e^+e^-\!\!$ colliders with different collision energies, assuming an integrated luminosity of
$\hs\mL\!=\hsm 5$\,ab$^{-1}${\hsm}.\ 
{As discussed in the text, the bounds on the conventional form factors (in blue color) 
are included for illustration only, since they are incompatible with the full SM gauge group with
spontaneous electroweak symmetry breaking, and hence are invalid.
}}
\label{tab:c}
\label{tab:4new}
\end{table}

\hspace*{1mm}
\subsection{\hspace*{-3mm}Analyzing Correlations between CPV nTGCs}
\label{sec:correlations}
\label{sec:3.2} 

In this subsection, we analyze the the correlations between each pair of the CPV nTGC form factors 
and between each pair of CPV dimension-8 nTGC operators.\ 

\begin{figure}[t]
\centering
\vspace*{-3mm}
\includegraphics[height=7cm]{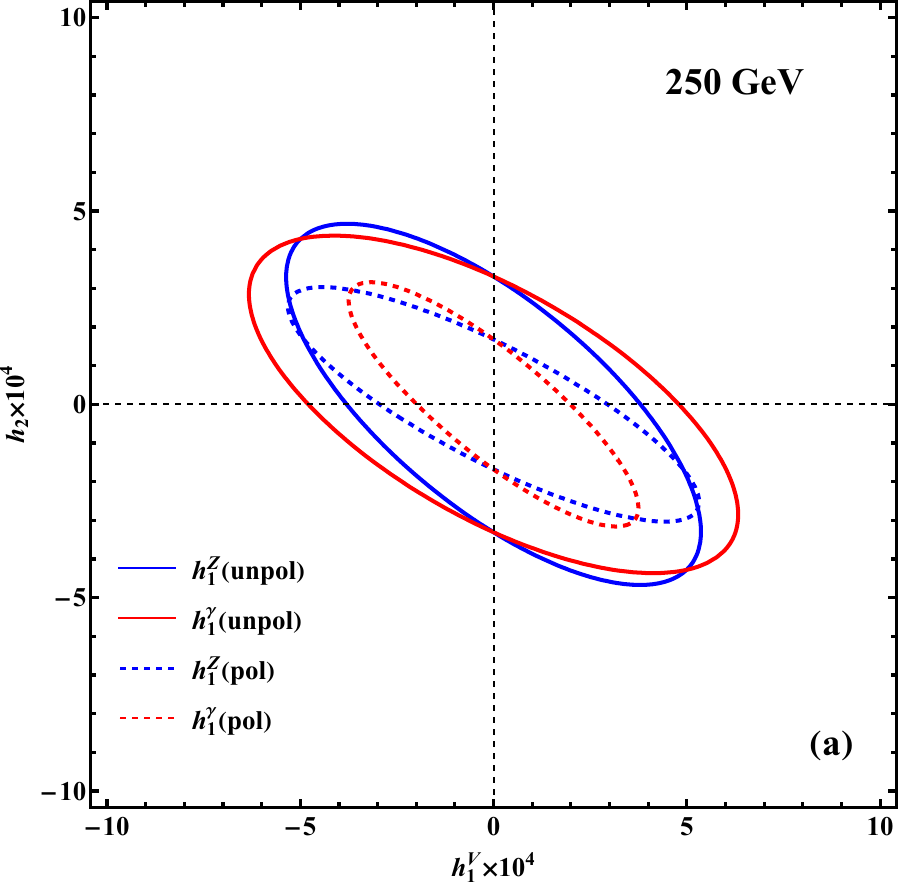} 
\includegraphics[height=7cm]{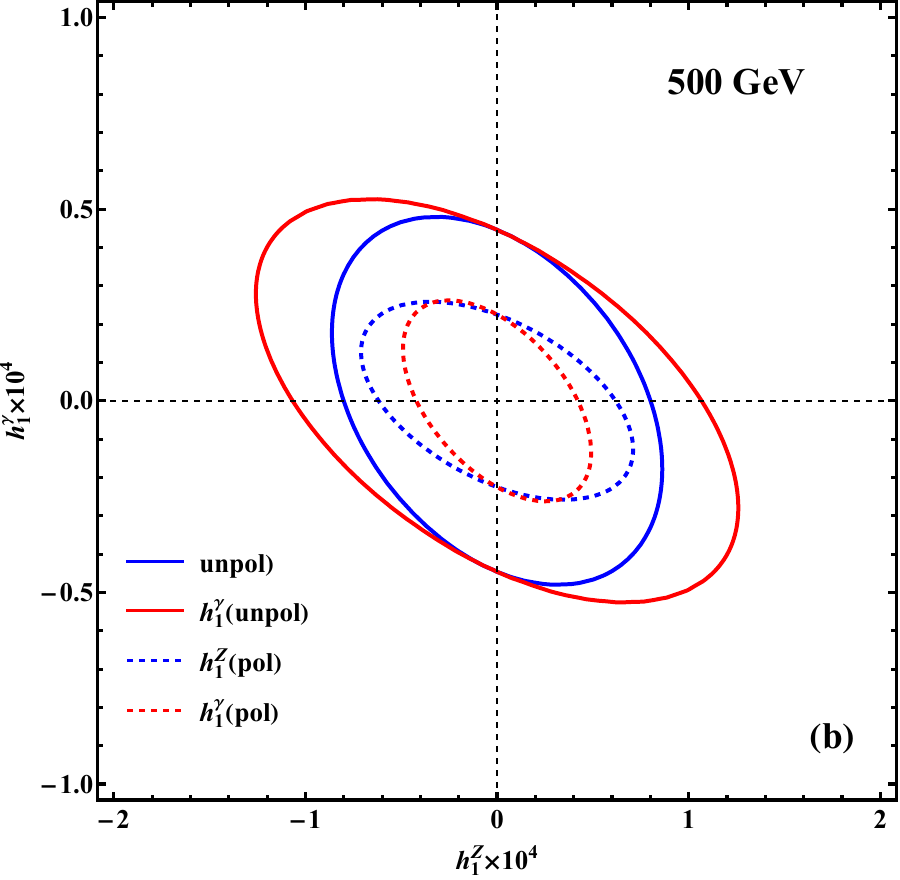} 
\\
\includegraphics[height=7cm]{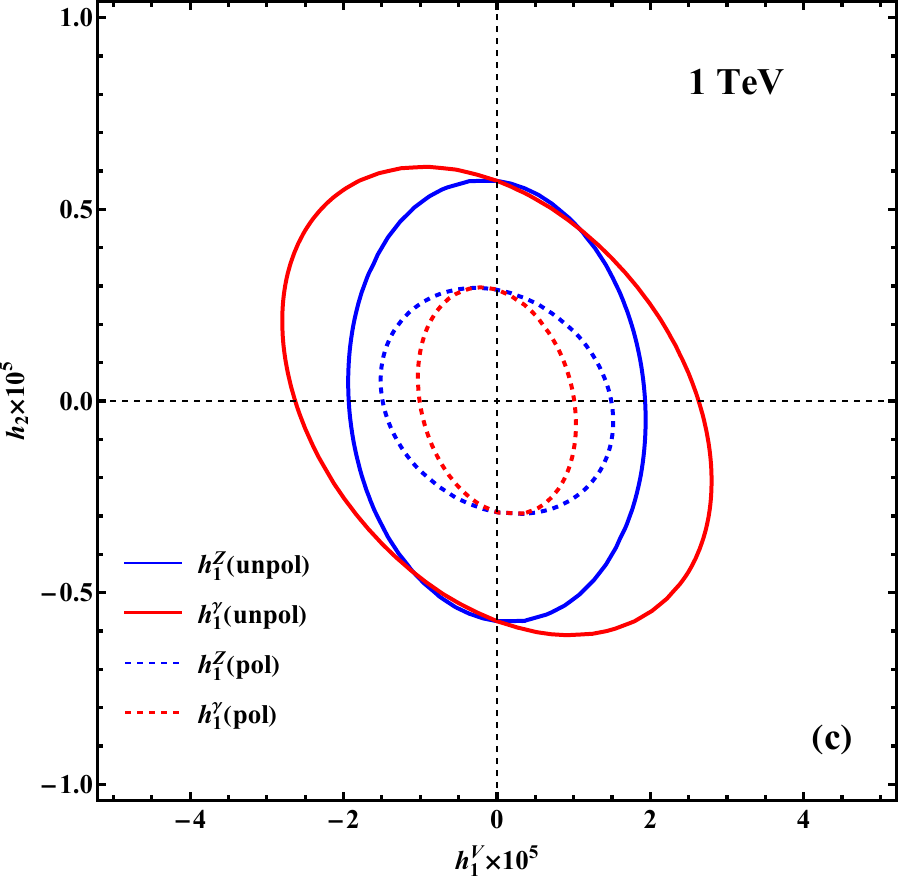} 
\includegraphics[height=7cm]{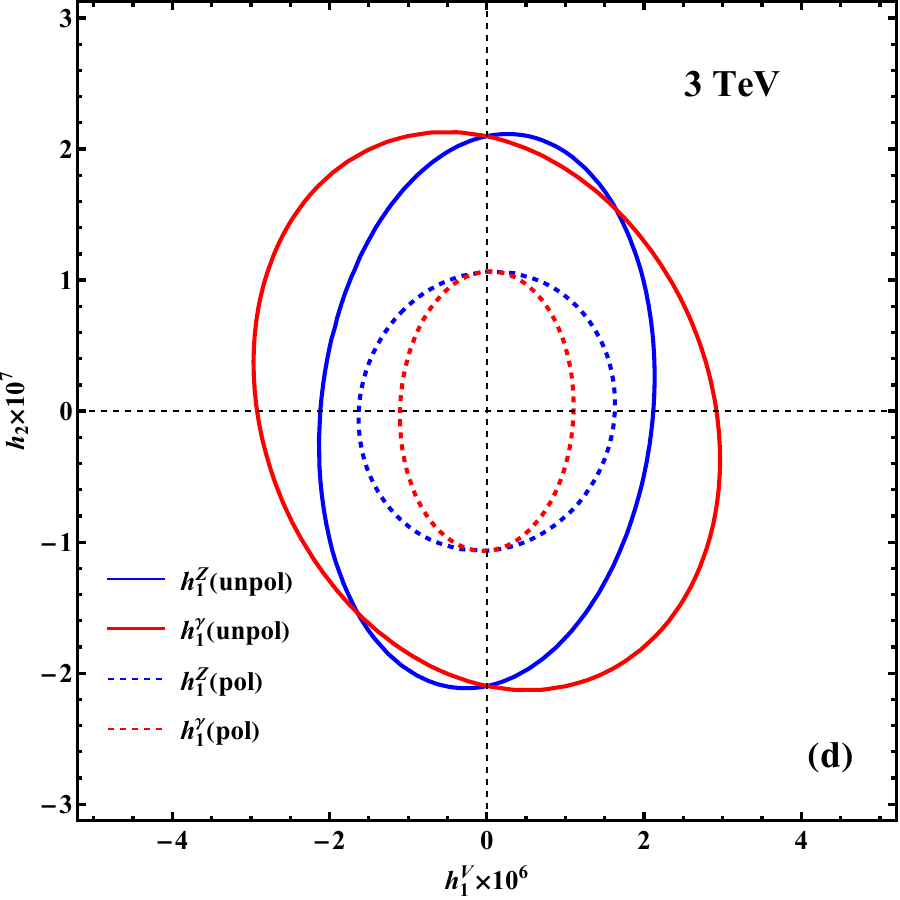} 
\hspace*{1.5mm}
\vspace*{-1mm}
\caption{\small\hspace*{-1mm}%
Correlation contours ($2\sigma$ bounds) for each pair of CPV nTGC form factors 
$(h_1^Z,h_2^{})$ and $(h_1^\ga,h_2)$ at 
$e^+e^-\!$ colliders for either unpolarized $e^\mp$ beams (blue color) or polarized $e^\mp$ beams (red color)  
where we choose $(P_L^e,\hs P_R^{\bar e})\hsm =(0.9,\hs 0.65)$.\ 
At each collison energy, we set a sample integrated luminosity of
$\hs 5$\,ab$^{-1}$.\  
The plots\,(a)-(d) present the correlation contours for the $e^+e^-$ collider energy
$\sqrt{s\,}\hsm =\hsm (0.25,\hs 0.5,\hs 1,\hs 3)\hs$TeV, respectively.
}
\label{fig:3}
\label{fig:4new}
\end{figure}

For this analysis, we first define $\chi^2$ likelihood functions as follows:
\beqs
\begin{align}
\chi^2_f &\,=\, {\sum_j\!\frac{\,S_j^2\,}{ B_j}}
= {\sum_j\!\frac{\,\sigma_{1,j}^2\,}{\si_{0,j}^{}}\times \mL\times\epsilon} \, ,
\\
\chi^2 &\,=\, {\sum_f \chi_f^2}\simeq {3\chi^2_d + 2\chi_u^2 + 3\chi_{\ell}^2} \, .
\end{align}
\eeqs
We present in Fig.\,\ref{fig:3} the $\chi^2$ contours for the correlations 
between each pair of the CPV nTGC form factors 
$(h_1^Z,\hs h_2^{})$ and $(h_1^\ga,\hs h_2^{})$ at the $2\sigma$ level.\  
We recall that in the high energy limit $s\!\gg\! M_Z^2$, 
the sensitivity reaches for the $h_1^V$ are determined by 
$\sin\phi_*$ and the sensitivity reaches for $h_2$ are determined by $\sin2\phi_*$; 
so $h_1^V$ and $h_2^{}$ are almost uncorrelated.\ However, when $s$ is not large, 
the $(h_1^V,\hs h_2^{})$ correlations are significant, 
and we see in Fig.\,\ref{fig:3}
how these correlations gradually diminish as the collision energy increases.\


%
\begin{figure}[]
\centering
\vspace*{-6mm}
\includegraphics[height=7cm]{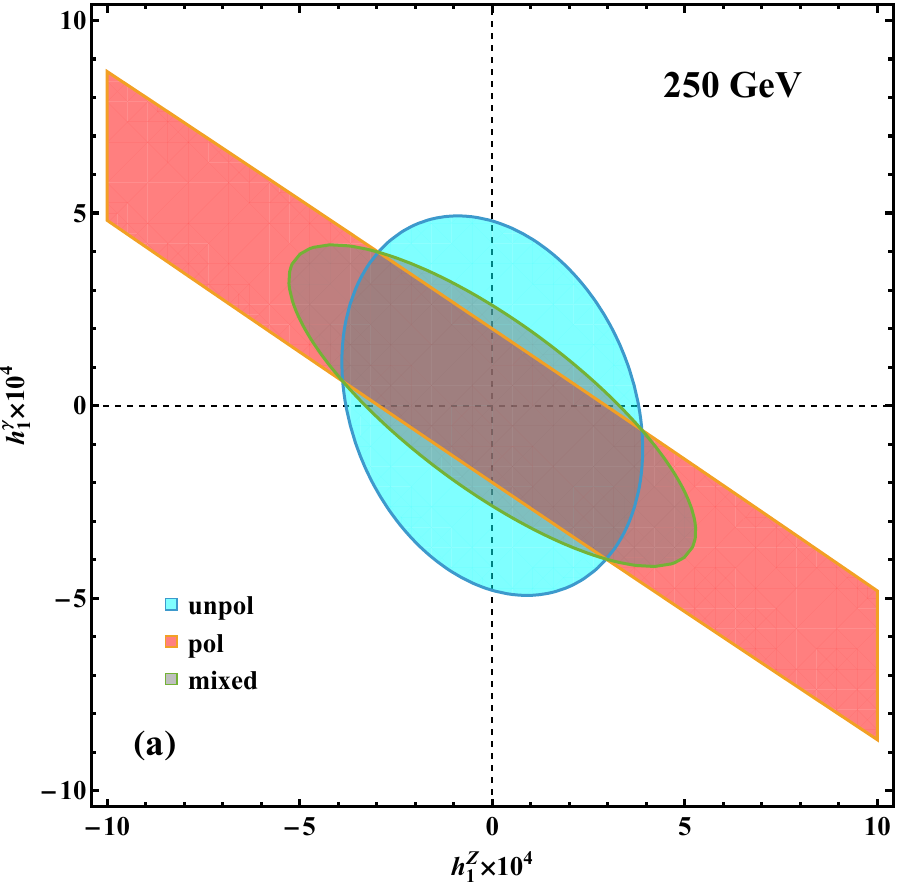}
\includegraphics[height=7cm]{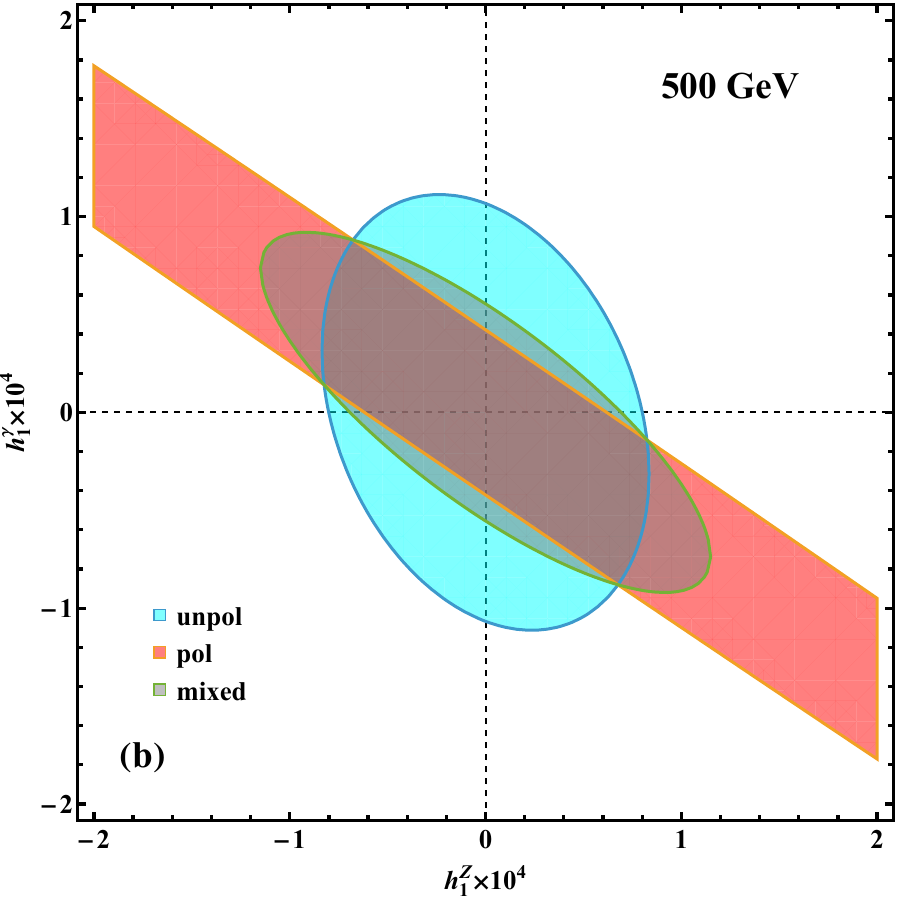}
\\
\includegraphics[height=7cm]{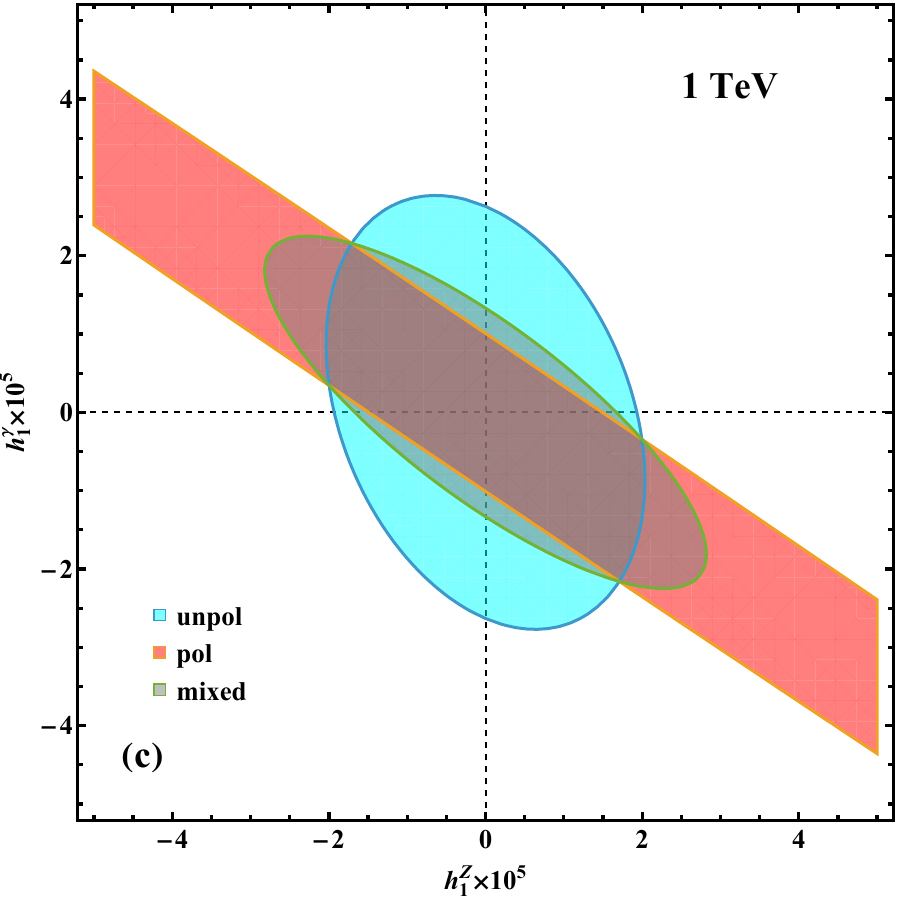}
\includegraphics[height=7cm]{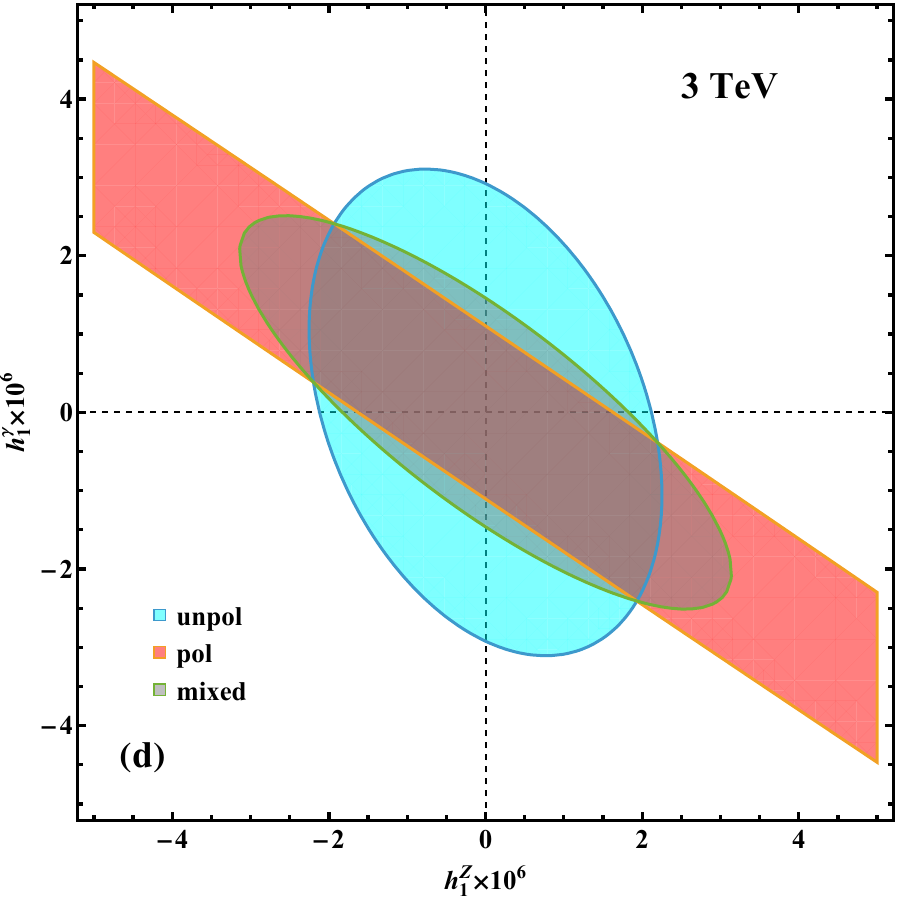}
\vspace*{-1mm}
\caption{\hspace*{0mm}\small 
Correlation contours ($2\sigma$ bounds) for each pair of the CPV nTGC form factors $(h_1^Z,\hs h_1^\ga)$ at $e^+e^-$ colliders 
with $\sqrt{s\,}\!=\!(0.25,\hs 0.5,\hs 1,\hs 3)\hs$TeV
and with an integrated luminosity of \,5\,ab$^{-1}$.\ 
In each plot, the correlation bound for the case of unpolarized
$e^\mp$ beams is shown as the blue contour and that for the case 
of polarized $e^\mp$ beams is given by the pink contour 
with the $e^{\mp}$ beam polarizations
$(P_L^e, P_R^{\bar{e}})\!=\!(0.9, 0.65)$;
whereas the light-blue contour presents the correlation bound for the mixed setup in which half of the data comes from unpolarized collisions and the other half of the data comes from polarized collisions.\  
The plots\,(a)-(d) present the correlation contours for the $e^+e^-$ collider energy
$\sqrt{s\,}\hsm =\hsm (0.25,\hs 0.5,\hs 1,\hs 3)\hs$TeV, respectively.
}
\label{fig:4}
\label{fig:5new}
\end{figure}

Figure\,\ref{fig:4} displays the correlations of 
the form-factor pair $(h_1^Z,\,h_1^\ga)$.\ 
Note that we can expand $c_{i}^{}(\theta,\,\theta_*^{})$ in Eq.(\ref{eq:ds1}) as follows:  
\begin{eqnarray}
c_i^{}(\theta,\theta_*) \,=\, 
P_L^eP_R^{\bar e}c_Lc_L^Vc_{i}^L(\theta,\theta_*)
+(1\!-\!P_L^e)(1\!-\!P_R^{\bar e}) c_R^{}c_R^Vc_{i}^R
(\theta,\theta_*) \,. 
\end{eqnarray} 
Thus, if $P_L^eP_R^{\bar e} \!\gg\!(1\!-\!P_L^e)(1\!-\!P_R^{\bar e})$,  
the contributions to both $h_1^Z$ and $h_1^\ga$ are dominated 
by the same term $c_{1}^L(\theta,\theta_*)$, so that $h_1^Z$ and $h_1^\ga$ 
become highly correlated in the polarized case.\
This is indeed the case since our choice 
$(P_L^e, P_R^{\bar{e}})\!=\!(0.9, 0.65)$ gives
$P_L^eP_R^{\bar e}\!\simeq\!0.59$
and $(1\!-\!P_L^e)(1\!-\!P_R^{\bar e})\!=\!0.035\hs$.\ 
This means that the beam polarizations can
tighten each correlation contour along the
minor axis of the ellipse as shown in Fig.\,\ref{fig:5new}.\ 
But this also makes the correlation bound much weaker
along the major axis of the ellipse
(orientated towards the northwestern direction
in the correlation plane).\
We explain this feature as follows.\
For purely left-handed or right-handed electrons
the $h_1^Z$ and $h_1^\gamma$ have the same angular distributions,
and thus their correlation contour should collapse into
a line with slope $-1\hs$.\
This means that for the polarized setup the correlation in each plot
is poorly constrained along the major axis of the ellipse
(orientated towards the northwestern direction in 
the $h_1^\gamma\hsmx -\hsm h_1^Z$ plane).\ 
Hence, this motivates us to modify the polarization setup 
and optimize the correlation bound along this poorly constrained 
direction by combining the data-taking from both the unpolarized 
collisions and polarized collisions.\
For this optimization, we present in each plot of Fig.\,\ref{fig:5new}
the correlation bound for a mixed setup
in which half of the data arises from the operation 
with unpolarized beams and the other half of the data arises 
from the operation with polarized beams, 
as shown by the light-blue contour.\
Fig.\,\ref{fig:5new} shows that the mixed setup can significantly 
enhance the correlation bound along the major axis of the ellipse 
and becomes comparable to that of the unpolarized setup.\ 
In addition, the correlation bounds in the polarized and mixed setups  
are much tighter than that of the unpolarized setup  
by about $(40\!-\!50)\%$ along the minor axis
of the ellipse contours.\

\hspace*{0.5mm}
\subsection{\hspace*{-3mm}Improvement from the Multivariable Analysis}
\label{sec:MVA}
\label{sec:3.3} 

%
\begin{figure}[t]
\centering
\includegraphics[height=5.5cm]{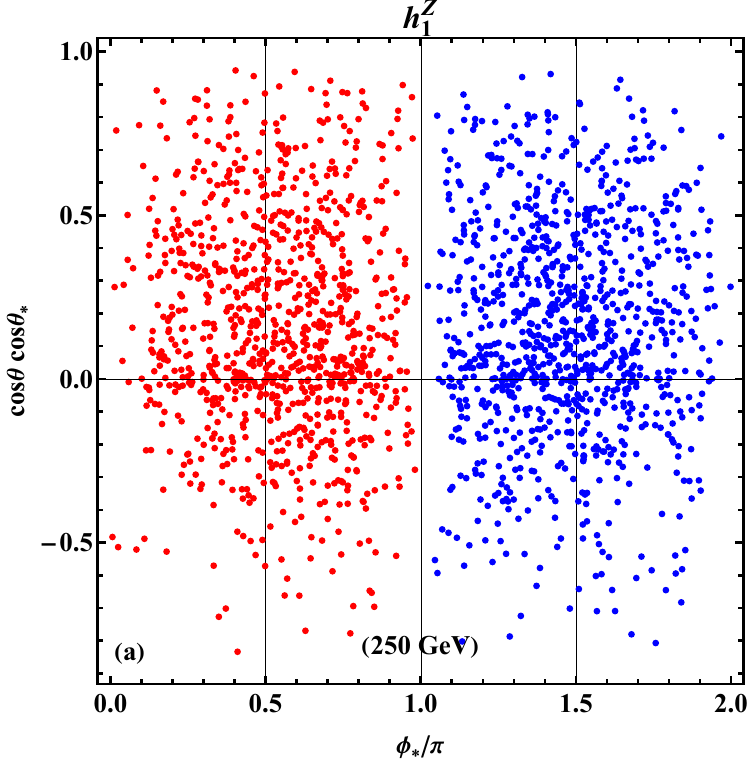}
\includegraphics[height=5.5cm]{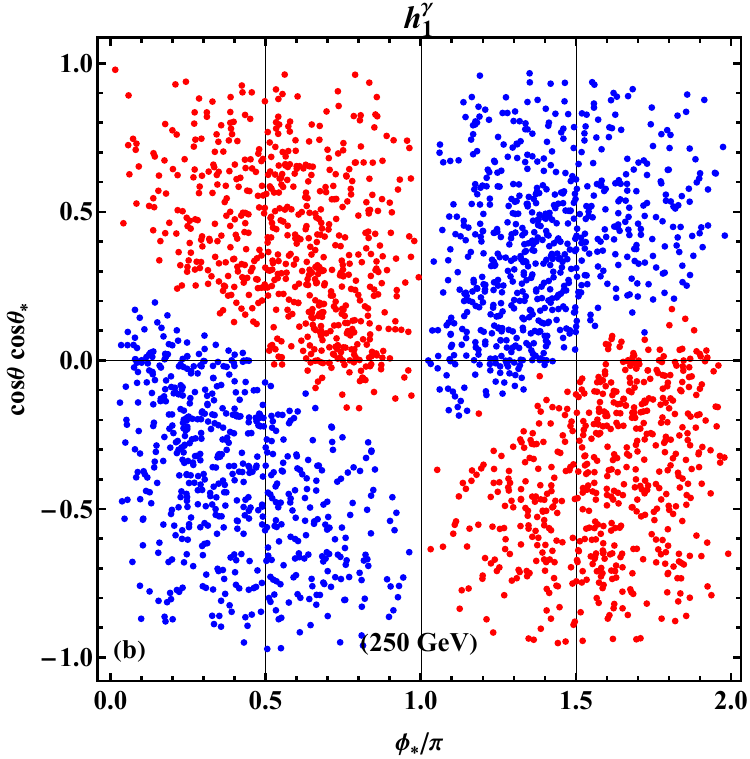}
\includegraphics[height=5.5cm]{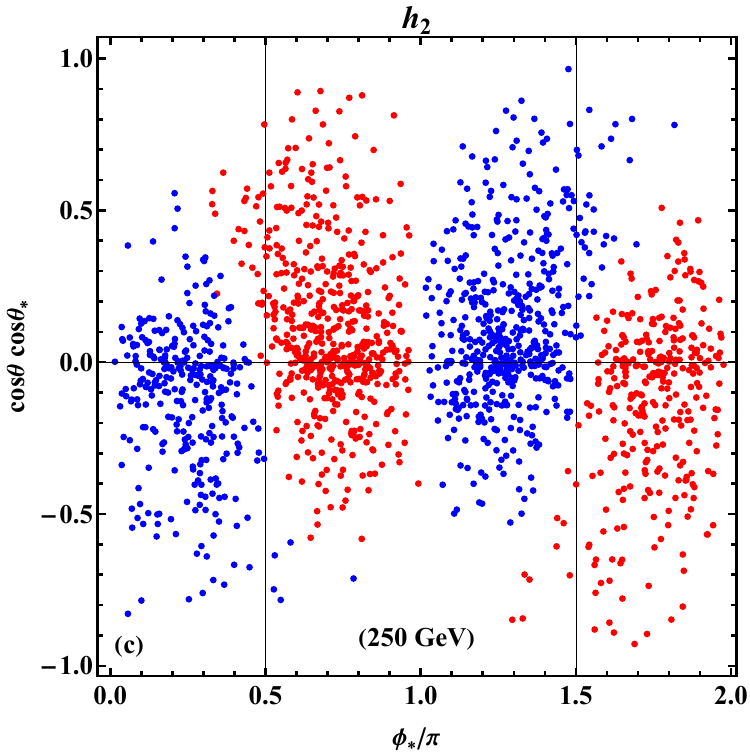}
\\
\includegraphics[height=5.5cm]{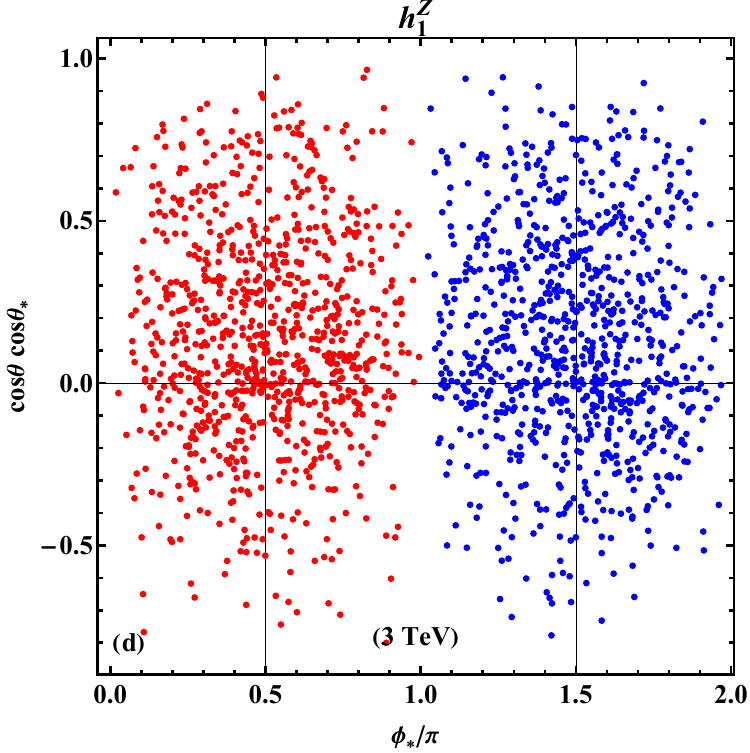}
\includegraphics[height=5.5cm]{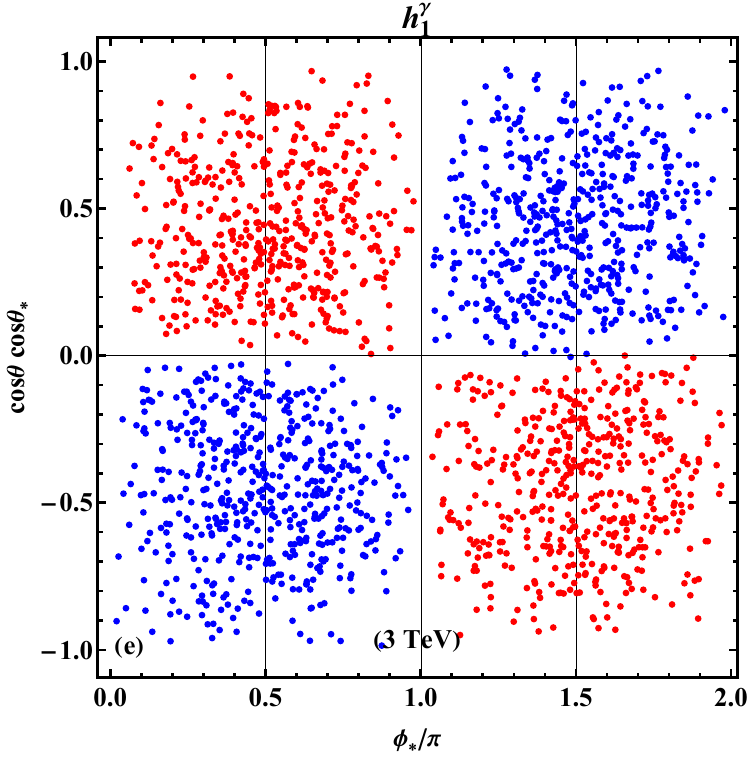}
\includegraphics[height=5.5cm]{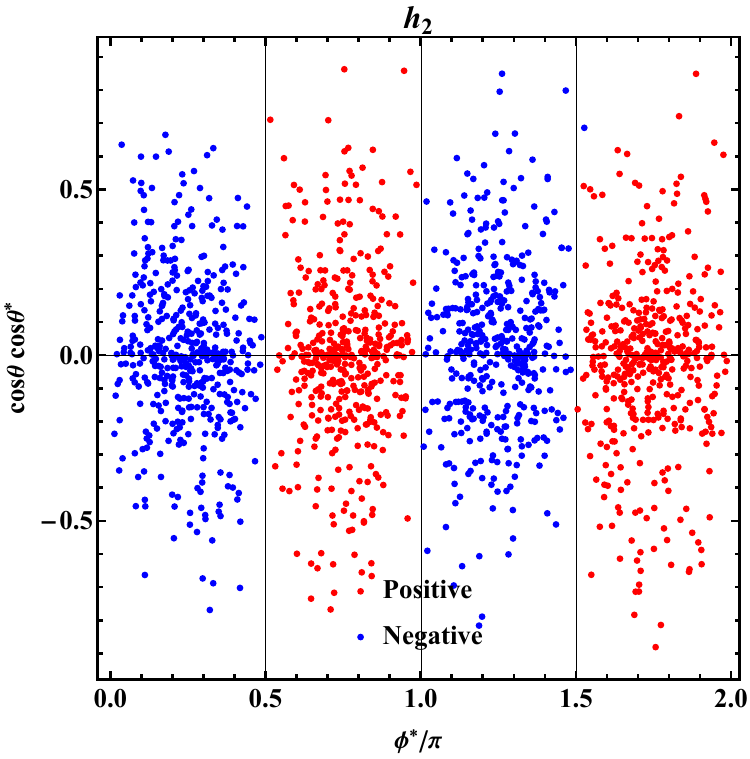}
\vspace*{-1mm}
\caption{%
		Distributions of the CPV nTGC form factors $(h_1^Z,\, h_1^\gamma,\,h_2^{})$ with simulated events 
		in the $(\phi^*\!,\, \cos\hsm\theta\cos\hsm\theta^*)$ plane for the reaction  
		$e^+e^-\!\ito Z\ga\ito \ell^-\ell^+\ga$  
		with collision energy $\sqrt{s\,}\!=\!250\,$GeV for plots\,(a)-(c) and
		$\sqrt{s\,}\!=\!3\,$TeV for plots\,(d)-(f).\ 
		In each plot, the red (blue) points present the case with positive (negative) interference term,
		and there are clear separations between the red-point events and the blue-point events.}
\label{fig:6}
\vspace*{3mm}
\end{figure}

In this subsection, we perform a multivariable analysis (MVA) to improve the sensitivity reaches on the CPV nTGCs.\ 

Inspecting the structure of the interference term (\ref{eq:ds1}), we see that the form of the interference term $\sigma_1^{}$ 
is determined by the three kinematic angles $\theta$, $\theta_*^{}$, and $\phi_*^{}\hs$.\  
Thus, under high energy expansion we can write down the asymptotic behavior of Eq.\,\eqref{eq:ds1} as follows: 
\begin{align}
\label{eq:ds1x}
\hspace*{-5mm} 
\frac{\d^3\sigma_1}{\,\d\theta \d\theta^{}_*\d\phi^{}_*}
=h_1^V\!\!\left[\frac{\sqrt{s\,}\,}{M_Z} P_1^{}(\cos\hsm\theta,\cos\hsm\theta^{}_*)\sin\phi^{}_*
\!+\hsm O\big({{s}^0}\big)\!\right] \!\!+\!
h_2\!\!\left[\hsm\frac{ s}{M_Z^2} P_2^{}(\cos\hsm\theta,\cos\hsm\theta^{}_*)\hsm\sin2\phi^{}_*
\!+\!O\big(\hsmx\sqrt{s\,}\hs\big)\!\right]\!,
 \end{align}
where the sign of functions $P_{1}$ and $P_2$ are determined by $\cos\hsm\theta\cos\hsm\theta^{}_*$.\ 
We note that the form factors $h_2^{}$ has leading energy power of $s^1$ with angular dependence of $\sin\hsm\phi^{}_*$,
whereas the form factors $h_1^{V}$ has leading energy power of $\sqrt{s\,}$ with angular dependence of $\sin\hsm 2\phi^{}_*$,

To separate the positive and negative contributions of the interference term, 
we can exploit their multidimensional distributions.\ To this end, we define a new variable 
$\omega \hsm\equiv\hsm \cos\hsm\theta \cos\hsm\theta^*$ \cite{Liu:2024tcz}.\ 
We present in Fig.\,\ref{fig:6} 
 the distributions of the CPV nTGC form factors $(h_1^Z,\, h_1^\gamma,\,h_2^{})$ with simulated events 
 in the $(\phi^{}_*,\, \omega)$ plane for the reaction  
 $e^+e^-\hsm\!\ito\hsm Z\ga$ (with $Z\!\ito\ell^-\ell^+$)  for collision energy $\sqrt{s\,}\!=\!250\,$GeV [as shown in plots\,(a)-(c)] and
 $\sqrt{s\,}\!=\!3\,$TeV [as shown in plots\,(d)-(f)].\ 
 In each plot, the red (blue) points present the case with positive (negative) interference term.\ 
%
Fig.\,\ref{fig:6} shows that positive-contribution signal events (red points) and negative-contribution signal events (blue points) 
are clearly separated in the $(\phi_*^{},\, \omega)$ plane.\ 
We have 8 regions which are divided by signs of $(\omega,\,\sin\phi^{}_*,\,\sin2\phi_*)$.\ 
For Fig.\,\ref{fig:6},  we find that for each form factor $h_1^V$ or $h_2^{}$ 
the contributions of the sub-leading terms in Eq.\eqref{eq:ds1x} can have different angular dependence on $\phi_*^{}$, 
$\theta$, and $\theta_*^{}$, which are still visible at $\sqrt{s\,} \!=\!250$\,GeV;  
whereas at $\sqrt{s\,} \!=\!3$\,TeV,  the contributions of the sub-leading terms in each case are negligible and the leading terms
in Eq.\eqref{eq:ds1x} dominate over the interference cross section.\ 
In consequence, for the much higher collision energy $\sqrt{s\,} \!\!=\hsm\!3\hs$TeV, 
the boundaries of the positive signal region and the negative signal region are more aligned with the lines of 
$\omega\hsm\! =\! 0$,  $\sin\hsm\phi^{}_*\hsm\!=\! 0$, and $\sin\hsm 2\phi_*^{}\hsm\!=\! 0\hs$.\
On the other hand, for the lower collision energy $\sqrt{s\,} \!\!=\hsm\!250\hs$GeV,
the boundaries of the positive signal region and the negative signal region become more irregular and
have visible deviations from the lines of 
$\omega\hsm\! =\! 0$,  $\sin\hsm\phi^{}_*\hsm\!=\! 0$, and $\sin\hsm 2\phi_*^{}\hsm\!=\! 0\hs$.\

\begin{table}[t]
\centering
\begin{tabular}{c||c|c|c}
\hline\hline
		&&\\[-3.5mm]				
		$\sqrt{s\,}$\,(TeV)\, 
		& $|h_2^{}|$(original,{\hs}improved) & $|h_1^Z|$(original,{\hs}improved) & 
		$|h_1^\gamma|$(original,{\hs}improved)
		\\
		&& \\[-3.7mm]
		\hline
		&& \\[-3.8mm]
		0.25   & (2.7,\,2.4)$\times10^{-4}$
		& (3.1,\,3.0)$\times10^{-4}$
		& (3.9,\,3.7)$\times10^{-4}$					
		\\
		\hline
		\hline
\end{tabular}
\vspace*{-0.5mm}
\caption{\small 
Sensitivity reaches ($2\hs\sigma$ bounds) for probing the 
CPV nTGV form factors through the reaction $\,e^-e^+\!\!\to\!Z\ga$ 
(with $Z\!\to\hsm  q\bar{q},\hs \ell\bar{\ell}\hs$)  				
at an \hs$e^+e^-\!\!$ collider with collision energy $\sqrt{s}\!=\!250\hs$GeV and integrated luminosity $\mL\! =\! 5\hs$ab$^{-1}$,
for the original and improved MVA methods.\  
The kinematic cut $|M_{f\bar f}^{}-\hsm M_Z^{}|\!<\!10$\,GeV 
is imposed and the MVA cuts are described in the text.}
\label{tab:4}
\label{tab:5new}		
\end{table}

Employing multivariable analysis (MVA), 
we have constructed a decision boundary in the $(\phi_*^{},\hs \omega)$ plane 
to distinguish effectively between events with positive and negative interference cross sections.\ 
We show in Table\,\ref{tab:4} a comparison of the sensitivity reaches for the original method and the improved MVA method.\ 
From Table\,\ref{tab:4}, we find that, since the boundaries of positive and negative contributions are very close to the boundaries 
of simple kinematic cuts on the angular variables, 
the MVA method only improves slightly the sensitivity reaches over the original method.

\hspace*{0.5mm}
\subsection{\hspace*{-3mm}Comparison of Sensitivities between Lepton and Hadron Colliders }
\label{sec:3.4}

\begin{table}[t]
\tabcolsep 1.5pt
\centering
\begin{tabular}{c|c||c|c|c}
\hline\hline
		&&&\\[-3.5mm]				
		$\sqrt{s\,}$\,(TeV)\, & ~$\mathcal L$\,(ab$^{\hsm -1}$)~
		& $|h_2|$ & $|h_1^Z|$ & $|h_1^\gamma|$
		\\
		&&& \\[-3.7mm]
		\hline\hline
		&&& \\[-3.8mm]
		~$e^+e^-${\hs}(0.25)~  & 5 & ({2.7},\,{1.4})$\times10^{-4}$
		& ({3.1},\,{2.4})$\times10^{-4}$
		& ({3.9},\,{1.6})$\times10^{-4}$					
		\\
		&&& \\[-3.8mm]
		\hline
		&&& \\[-3.8mm]
		~$e^+e^-${\hs}(0.5)~ &5 &	({3.6},\,{1.8})$\times10^{-5}$
		& ({6.5},\,{5.1})$\times10^{-5}$ & ({8.8},\,{3.4})$\times10^{-5}$
		\\
		&&& \\[-3.8mm]
		\hline
		&&& \\[-3.8mm]
		~$e^+e^-${\hs}(1)~  &5  &	({4.7},\,{2.4})$\times10^{-6}$
		& ({1.6},\,{1.2})$\times10^{-5}$& ({2.1},\,{0.82})$\times10^{-5}$
		\\
		&&& \\[-3.8mm]
		\hline
		&&& \\[-3.8mm]
		~$e^+e^-${\hs}(3)~  &5 & ({1.7},\,{0.87})$\times10^{-7}$
		&({1.7},\,{1.3})$\times10^{-6}$ & ({2.4},\,{0.90})$\times10^{-6}$
		\\
		&&&\\[-3.8mm]
		\hline
		&&& \\[-3.8mm]
		~$e^+e^-${\hs}(5)~  &5 &\,	({3.7},\,{1.9})$\times10^{-8}$
		\,&\, ({6.2},\,{4.8})$\times10^{-7}$
		\,&\, ({8.6},\,{3.3})$\times10^{-7}$ \,
		\\
		\hline\hline
		&&&& \\[-4mm]
		& 0.14 & \,9.6$\times10^{-6}$\,
		& \,1.9$\times10^{-4}$\,&\,2.2$\times10^{-4}$\,\\
		&&&& \\[-4.4mm]
		LHC{\hs}(13) &0.3& \,7.5$\times10^{-6}$\, &\,1.5$\times10^{-4}$\,&\,1.8$\times10^{-4}$\,\\
		&&&& \\[-4.4mm]
		& 3 & \,3.8$\times10^{-6}$\, &\,0.80$\times10^{-4}$\,&\,0.97$\times10^{-4}$\,\\
		\hline
		&&&& \\[-4mm]
		&3& \,4.0$\times10^{-9}$\, &\,6.1$\times10^{-7}$\,&\,7.2$\times10^{-7}$\,\\
		&&&& \\[-4.4mm]
		$pp${\hs}(100) & 10 & \,2.6$\times10^{-9}$\,  &\,4.2$\times10^{-7}$\,&4.9$\times10^{-7}$\\
		&&&& \\[-4.4mm]
		& 30 & \,1.9$\times10^{-9}$\, &\,3.0$\times10^{-7}$\,&\,3.5$\times10^{-7}$\,\\
		\hline\hline
	\end{tabular}
\vspace*{0mm}
\caption{{\small\hspace*{-1mm}
			Sensitivity reaches ($2\hs\sigma$ bounds) on the nTGC form factors at 
			of ${\hs}e^+e^-\!$ colliders with different collision energies, in comparison with those of the LHC}
		and $pp$\,(100\,TeV)$\!$ colliders.\
		The reactions $\,e^-e^+\!\!\to\!Z\gamma$ (with $Z\!\!\to\!\! q\bar{q}$)
		and $\,pp(q{\hs}\bar{q})\!\!\to\!\!Z\gamma$ (with $Z\hsm\!\!\to\!\!\ell\bar{\ell},\hs\nu\bar{\nu}\hs$)
		are considered for the lepton and hadron colliders respectively.\ 
		For the $e^+e^-$ colliders, each entry corresponds to
		(unpolarized,\,polarized) $e^\mp$ beams, and 
		we choose the benchmark inputs for the $e^\mp$ beam polarizations 
		$(P_L^e, P_R^{\bar{e}})\!=\hsm (0.9,\hs 0.65)$.}
\label{tab:6new}
\end{table}

In this subsection, we compare the sensitivity bounds on probing the nTGCs at $e^+e^-$ colliders
versus hadron colliders.\ 
We note that the hadron colliders (such as the LHC and future high-energy $pp$ colliders) 
generally have higher collision energy but  
lower luminosity than the lepton colliders (such as the CEPC, FCC-ee, ILC and CLIC).\ 
Since the squared term of nTGC has higher energy-power dependence than the interference term, 
so its contribution dominates the cross section over the interference term for $pp$ colliders.\ 
Moreover the contribution of a CPV nTGC to the scattering amplitude 
differs from that of the related CPC nTGC by an overall factor of the imaginary number $\ii\hs$
(apart from simple normalization factors of the CPC versus CPV nTGCs).\ 
Hence, at hadron colliders the contributions of the CPC and CPV nTGCs 
to the cross section are dominated by their squared terms respectively, 
and thus they are mainly the same 
(apart from simple normalization factors of the CPC versus CPV nTGCs)\,\cite{Ellis:2023ucy}.\ 

\vspace*{0.6mm}

\begin{figure}[t]
\centering
\includegraphics[width=8.2cm,height=6cm]{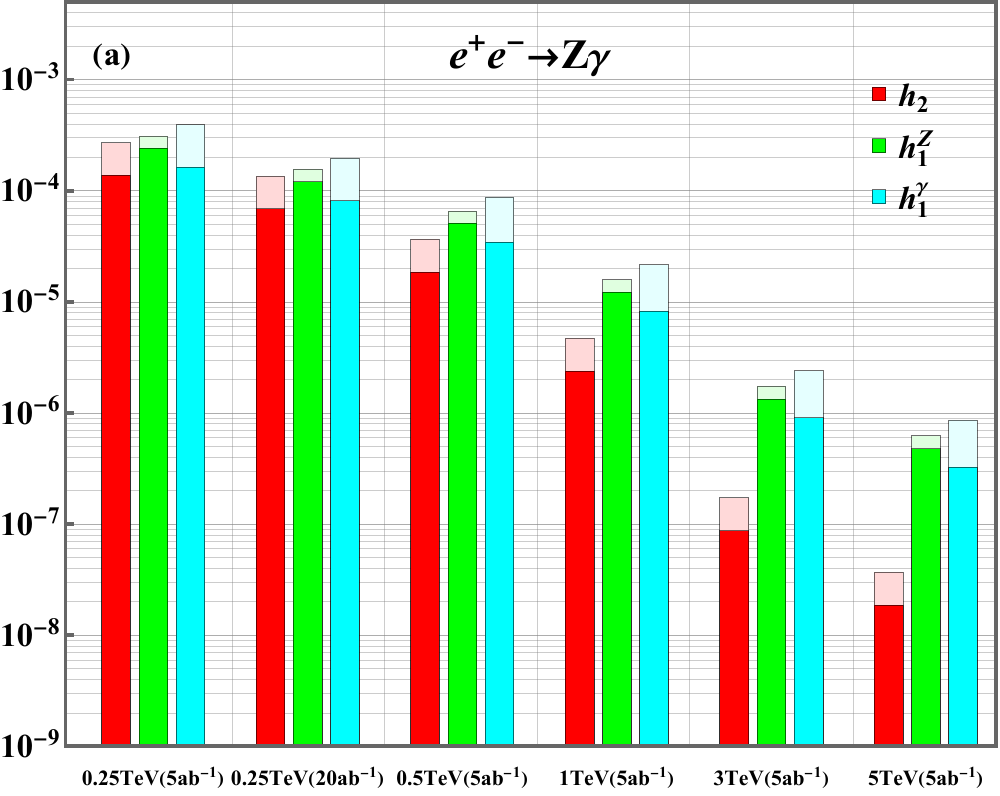}
\includegraphics[width=8.2cm,height=6cm]{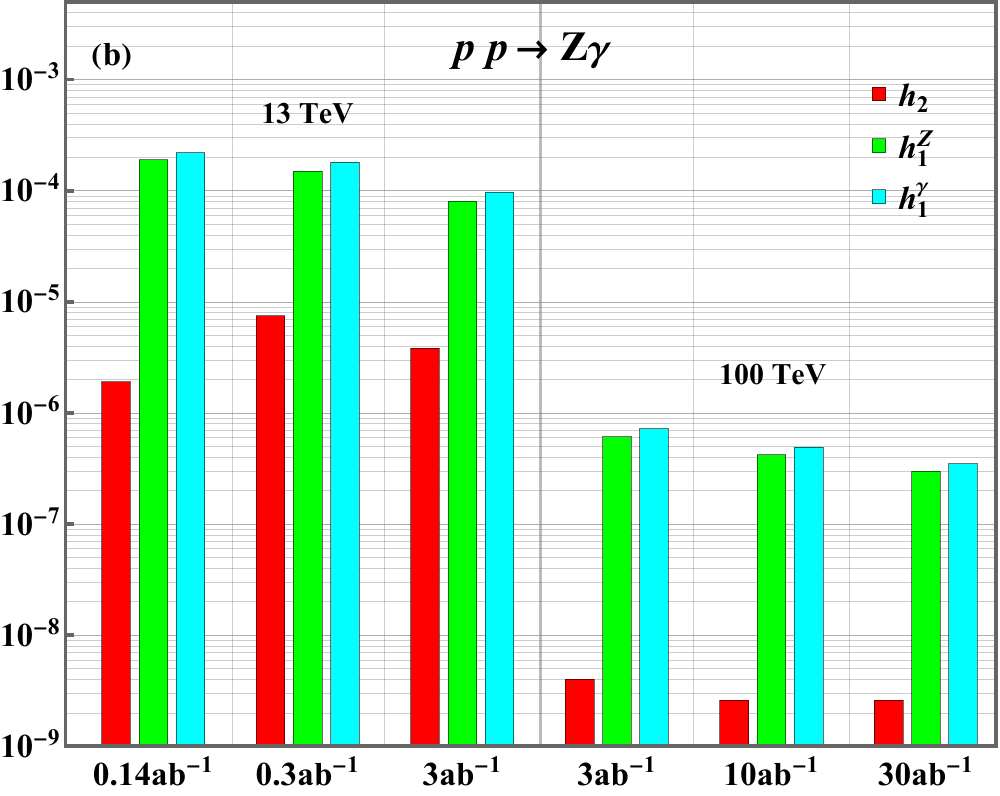}
\vspace*{0mm}
\caption{\small
Comparison of sensitivity reaches ($2\hs\sigma$ bounds) on the nTGC form factors 
at \hs$e^+e^-\!\!$ colliders [left plot\,(a)] and hadron colliders [right plot\,(b)].\
In plot\,(a) the results are shown for the reaction $\,e^-e^+\!\!\to\!Z\gamma$ (with $Z\!\to\! q\bar{q}$)
at different collision energies $\sqrt{s\,}\!=\!(0.25,\hs 0.5,\hs 1,\hs 3,\hs 5)\hs$TeV, whereas in plot\,(b) 
the results are shown for the LHC and $pp$\,(100\,TeV)$\!$ colliders with different integrated luminosities 
via the reaction $\,pp(q{\hs}\bar{q})\hsm\!\to\!Z\gamma$ (with $Z\!\to\!\ell\bar{\ell},\hs\nu\bar{\nu}\hs$).\
For the $e^+e^-$ colliders in plot\,(a), the sensitivity reaches with light (heavy) color in each entry 
correspond to unpolarized (polarized) $e^\mp$ beams, and  
we choose the benchmark inputs for the $e^\mp$ beam polarizations 
$(P_L^e, P_R^{\bar e})=(0.9,\hs  0.65)$.}
\label{fig:7}
\vspace*{3mm}
\end{figure}

Then, we make use of our previous nTGC analyses 
for hadron colliders\,\cite{Ellis:2022zdw}\cite{Ellis:2023ucy}, 
and present a comparison of the sensitivity bounds in Table\,\ref{tab:6new}  and Fig.\,\ref{fig:7}
for probing the CPV nTGC faorm factors $(h_2^{},h_1^V)$ 
at the high energy $e^+e^-$ colliders versus $pp$ colliders.\ 
For instance, we find that the sensitivity bounds on the CPV nTGC factor $h_2^{}$  at the LHC 
are close to the corresponding bounds on  $h_2^{}$ at a 1\,TeV $e^-e^+$ collider, 
whereas the sensitivity bounds on $h_1^V$ at the LHC are close to 
the corresponding bounds at a 250\,GeV $e^+e^-$ collider.\ 
The high energy $pp\hs$(100\,TeV) colliders can provide the best sensitivity reaches 
on probing the CPV nTGCs.

\section{\hspace*{-2.5mm}Conclusions}
\label{sec:conx}
\label{sec:4}

The origin of the cosmological baryon asymmetry requires CP violation
beyond\,\cite{BA} the Kobayashi-Maskawa mechanism\,\cite{KM} in the SM, and we have shown
in this work that the neutral triple gauge couplings (nTGCs) generated by 
dimension-8 SMEFT operators offer new opportunities to search for additional
sources of CP violation.\ As discussed in Section\,\ref{sec:CPVnTGCs}, there are
3 relevant CPV dimension-8 operators in Eq.(\ref{eq:dim8H}) involving Higgs fields, 
and we have identified 2 additional CPV dimension-8 operators (\ref{eq:OG+G-}) 
that do not involve Higgs fields.\ The corresponding CPV neutral triple gauge vertices
(nTGVs) are given in Eqs.(\ref{eq:ZAV-dim8WB}) and (\ref{eq:ZAV-dim8G+-}), 
and their relations to the corresponding nTGC form factors that are compatible with 
the full electroweak gauge group 
SU(2)$_{\rm L}^{}\otimes\hs$U(1)$_{\rm Y}^{}$ with spontaneous symmetry breaking 
are given in Eq.(\ref{eq:ZAV*-FFnew2}).

\vspace*{0.6mm}

The sensitivity reaches on these CPV nTGV form factors and on the cutoff scales of the corresponding CPV nTGC operators  
were presented in Fig.\,\ref{fig:1} and Tables\,\ref{tab:f-p} and \ref{tab:o-p}, respectively.\ 
We performed the analysis for $e^+e^-$
(or $\mu^+\mu^-$\!) colliders with collision energies 
$\sqrt{s\,}\!=\!(0.25,\hs 0.5,\hs 1,\hs 3,\hs 5)\hs$TeV, choosing  
integrated luminosities of 5\,ab$^{-1}$ in each case and considering both unpolarized and
polarized $e^\mp$ beams.\  The sensitivity reaches on probing the cutoff scales of the CPV dimension-8
SMEFT operators such as $\OO_{\!\tilde{G}_+}^{}\!\!\!$ 
were found to range from about 1\,TeV at $\sqrt{s\,} \!=\!250$\,GeV to
about $(8\hsm -\hsm 9)\hs$TeV at $\sqrt{s\,} \!=\!3$\,TeV, and   
to over 10\,TeV at $\sqrt{s\,} \!=\!5$\,TeV, as shown in Table\,\ref{tab:3new}.\ 
On the other hand, we estimated the sensitivity reaches on the CPV nTGC form factors and found 
that the sensitivity reaches on these form factors vary from 
${O}(10^{-4})$ to ${O}(10^{-6}\hsm\!-\hsm\!10^{-8})$ for the $e^+e^-$ collision energy from 250\,GeV to $(3\!-\!5)\hs$TeV,
as shown in Table\,\ref{tab:2new}.\  
We also demonstrated in Table\,\ref{tab:4new} that using the conventional CPV nTGC form factor formulas (which are
incompatible with spontaneous electroweak gauge symmtry breaking of the SM) lead to fake sensitivities
to the nTGCs which are over-strong by a factor of $2\!-\!3$ for the $e^+e^-$ collision energy $\sqrt{s\,}\!=\!250\hs$GeV,
by a factor of $O(10)$ for $\sqrt{s\,}\!=\!(0.5\!-\!1)\hs$TeV, and
by a factor of $O(10^2)$ for $\sqrt{s\,}\!=\!(3\!-\!5)\hs$TeV.\ 
Hence it is important to use our present new formulation of the CPV nTGC form factors
that is fully consistent with the spontaneous electroweak gauge symmtry breaking.\   
We have further studied the correlations between
the sensitivities to pairs of CPV form factors, as shown in Figs.\,\ref{fig:4new} and \ref{fig:5new}.\ 
An exploratory multivariable analysis (MVA) was found to provide a minor
improvement in the sensitivities to the CPV nTGV form factors.\ 
We have further presented a comparison of the sensitivity bounds in Table\,\ref{tab:6new}  and Fig.\,\ref{fig:7}
for probing the CPV nTGC faorm factors $(h_2^{},h_1^Z,h_1^{\ga})$ 
at the high energy $e^+e^-$ colliders versus $pp$ colliders.\

\vspace*{0.6mm}

Our results demonstrate the interest in searching for CP-violating nTGCs in
$e^+ e^-$ (or, $\mu^+ \mu^-$) collisions at center-of-mass energies of 250\,GeV
and beyond.\ As we have shown, the prospective sensitivity reaches on probing the CPV SMEFT operator
scales extend significantly beyond the center-of-mass energies considered.\
Since these operators are presumably generated by certain high-energy 
ultraviolet-complete extension of the SM, studies of CP-violating nTGCs could
offer valuable insights to understanding its dynamics.

\vspace*{8mm}
\noindent 
{\bf\Large Acknowledgements}
\\
The work of J.E.\ was supported in part by the United Kingdom STFC Grant ST/T000759/1.\
The works of HJH and RQX were supported in part 
by the National Natural Science Foundation of China (NSFC)
under Grants 12175136 and 12435005, 
by the Shenzhen Science and Technology Program (under Grant JCYJ20240813150911015), 
by the State Key Laboratory of Dark Matter Physics,
by the Key Laboratory for Particle Astrophysics and Cosmology (MOE), and 
by the Shanghai Key Laboratory for Particle Physics and Cosmology.\ 
RQX is also supported in part by the National Science Foundation of China under Grants 
Nos.\,12175006, 12188102, 12061141002, by the Ministry of Science and Technology 
of China under Grants No.\,2023YFA1605800, and by the state key laboratory of nuclear physics and technology.




\end{document}